%% file: Danescu_Aperiodic.tex
\documentclass[preprint,5p,twocolumn,10pt]{elsarticle}
\usepackage{amssymb,amsmath,mathtools,amsthm,amsbsy}
\usepackage{cuted,graphicx,array,cuted}
\usepackage[format=hang]{caption}
\usepackage[rgb]{xcolor}
\usepackage{tikz,pgfplots}
\usepackage{afterpage}
\newcommand{\bs}{\boldsymbol}
\pgfplotsset{compat=1.18,/pgf/number format/.cd,fixed}
\begin{document}

\begin{frontmatter}

\title{Macroscopic elasticity of the hat aperiodic tiling}
\author[label1,label2]{Romain Rieger}
\affiliation[label1]{organization={Université de Lyon, Lyon, VetAgro Sup, UPSP ICE 2021},
             addressline={A104, 1 Av. Bourgelat},
             city={Marcy l'Etoile},
             postcode={69280},
             country={France}}
\affiliation[label2]{organisation = {Univ. Lyon, École Centrale de Lyon, , 69134 Ecully, France},%Department and Organization
            addressline={36 Av. Guy de Collongue}, 
            city={Ecully},
            postcode={69134},
            country={France}}
\author[label3]{Alexandre Danescu\corref{cor1}}
\ead{alexandre.danescu@ec-lyon.fr}
\cortext[cor1]{Corresponding author}
\affiliation[label3]{organization={Lyon Institute of Nanotechnologies, Université de Lyon, CNRS - UMR 5270, Ecole Centrale de Lyon,\\ INSA Lyon, UCBL, CPE Lyon},
             addressline={36 Guy de Collongue},
             city={Ecully},
             postcode={69134},
             country={France}}
%====================================================
% ABSTRACT
%====================================================
\begin{abstract}

In this paper, we explore the macroscopic elastic behavior of the aperiodic but hyperuniform {\em ein stein} (single tile) tiling, as recently reported in \cite{Smith-Aperiodic-2023, Smith-Chiral-2023}. The first step involves assigning mechanical properties to the geometric pattern. The simplest approach includes near-neighbor type (NN) interactions (springs) along the edges of the geometric pattern. To eliminate zero modes, we also incorporate angular interactions using the simplest quadratic approximation—the Kirkwood-Keating three-body potential. We compute the macroscopic elastic response on circular domains for various realizations across an increasing sequence of length scales. As the domain radius increases, the set average at a fixed length scale of the obtained macroscopic Hooke tensor approaches the hyperplane of two-dimensional isotropic Hooke tensors. In a closer-to-experiment scenario, we also discuss a second mechanical setting where edges of all polygons in the considered domain are treated as one-dimensional continua (straight beams). These continua are endowed with both extension/compression, shear and bending energy within the framework of the Timoshenko theory. The macroscopic response of the hat tiling aligns with that of an isotropic elastic continuum.
\end{abstract}

%====================================================
% GRAPHICAL ABSTRACT
%====================================================
%\begin{graphicalabstract}
%\includegraphics[scale=0.7]{Graphic_abstract.png}
%\end{graphicalabstract}
%====================================================
% HIGHLIGHTS
%====================================================
%\begin{highlights}
%\item We endow the edges of the the recently discover {\em ein stein} aperiodic tilling of the plane with %mechanical interactions and we explore the macrosocpic properties of the resulting discrete continuum at the %macro-scale. Since the aperiodic tilling with a single tile is hyperuniform by construction, we conjecture %that the macroscopic material behavior is that of an isotropic continuum. 

%\item We implement two different mechanical interactions : a first discrete one, based on near-neighbor type %interactions and, in order to prevent zero-modes, angular interactions in the Kirkwood-Keating approximation. %Closer to an experimental setting, we also study the macroscopic behaviour when the aperiodic tilling is %endowed with the classical extension/compression, shear and bending elastic energies along all of its edges. %In both cases the numerical experiments performed on different realization at various length-scales indicate %that, as expected, the macroscopic elastic law tends to that of an isotopic material.    
%\end{highlights}

%====================================================
% KEYWORDS
%====================================================
\begin{keyword}
macroscopic elasticity, hyperunifom materials, aperiodic tiling of the plane
\PACS: 46.25.-y \sep 62.20.de
\MSC: 784B05 \sep 74E05 \sep 74E10
\end{keyword}
\end{frontmatter}

%====================================================
% Main manuscript
%====================================================
\section{Introduction}
\label{introduction}

Answering a long-standing open question, an aperiodic tiling of the plane that use a single tile was recently discovered \cite{Smith-Aperiodic-2023, Smith-Chiral-2023}. Previous aperiodic tilings of the plane using multi-tiles already known are commonly related to quasicrystals and well understood as particular slices in higher-dimensional periodic lattices. Recent results \cite{Socolar-2023} show that the hat tiling is also quasicrystalline, but posses additional symmetry properties reminiscent from the hexagonal lattice on which the tiling is build on. 

In order to investigate the physical properties of the hat tiling, one has to endow the tiling geometry with some minimal physical attributes. In the frame of condensed-matter physics, the most common choice \cite{Kohmoto-1986, Kohmoto-1986-2} is the vertex tight-binding model on the quasi-lattice \cite{Schirmann-2023}, where hopping between near-neighbor vertices has equal probabilities. As the hat tiling edges form a subset of the symmetry lines of the hexagonal lattice in the plane, the spectral and transport properties obtained using the above assumption inherit some of the main features of graphene, i.e., Dirac cones and six-fold symmetry. However, the practical realization of such structures and experimental evidence of the theoretical predictions require complex phases in order to reproduce the underlying physical characteristics of the system. 

In contrast to condensed-matter physical models, the macroscopic mechanical properties of a planar tiling are much easier to interpret: they reflect the material response at a length-scale which is much larger than that of the particle interactions. The main theoretical tool to achive this is the homogenization theory\cite{Jikov-1994, Braides-2002} and a qualitative homogenization  result (existence of the macroscopic response) for functionals that depend on the Penrose tiling geometry through the shape and orientation is provided in \cite{Braides-2009}. The fact that the Penrose tiling is a quasicrystal, i.e. a two-dimensional projection of a particular slice in a ${\mathbb R}^5$-lattice is one of the main ingredients of the proof. 
 
In an attempt to extend the class of classical long-range order of lattice interactions to include locally disordered materials, Stillinger and Torquato introduced the concept of hyperuniformity \cite{Torquato-2003, Torquato-2018}. The definition of hyperuniformity concerns the scaling law for the variance of the number of points in a spherical observation window. Hyperuniform materials  include disordered sphere packings and quasicrystals. Obviously, this purely geometric definition tacitly assumes that the physical nature of particle interactions is completely determined by the particles spatial distribution, a property that holds true in most physical systems (condensed matter and/or granular materials) but metamaterials with purposely tailored properties can deviate from this general rule. Notice that, as the hat tiling is built using a single tile, the scaling law for the hat tiling is obviously the same as for classical lattices, which are known to be hyperuniform.   

The interplay between the geometry of a tiling and the associated physical properties is, in general, an intricate question. It is clear that, with minor exceptions, some kind of uniformity in both geometric and physical characteristics is needed to obtain an isotropic material at the macroscale. One notable exception is the symmetry of the geometric pattern combined with the symmetry of the physical structure associated with it. In the case of the Penrose tiling\footnote{Here, Penrose tiling refers to the \textit{kite and dart} construction \cite{Gardner-1997}, also known as the $P2$ construction.}, which has local 5-fold (pentagonal) symmetry \cite{Steinhardt-1996}, by endowing it with a mechanical structure that also respects the 5-fold symmetry, one can expect to prove isotropy at the macroscopic scale by using symmetry arguments \cite{Danescu-1997}.

\begin{figure}[h!]
\centering
\begin{tikzpicture}[scale = 0.6]
\draw[line width = 0.5mm,fill = lightgray] (0, 0) -- (3/2, -0.8660) -- (3, 0) -- (5/2, 0.8660) -- (3/2, 0.8660) -- (3/2, 3*0.8660) -- (0, 4*0.8660) -- (-1/2, 3*0.8660) -- (-3/2, 3*0.8660) -- (-3/2, 0.8660) -- (-3, 0) -- (-5/2, -0.8660) -- (-1/2, -0.8660) --cycle;
\draw[red,line width = 0.5mm,fill = lightgray] (0, 0) -- (3/2, -0.8660) -- (3, 0) -- (5/2, 0.8660) -- (3/2, 0.8660) -- (3/2, 3*0.8660) -- (0, 4*0.8660) -- (-1/2, 3*0.8660) -- (-3/2, 3*0.8660) -- (-3/2, 0.8660) -- (-3, 0) -- (-5/2, -0.8660) -- (-1/2, -0.8660) --cycle;
\draw[black,thin,dotted] (3/2, -0.8660)  -- (5/2, 0.8660);
\draw[black,thin] (3/2, -0.8660)  -- (3/2, 0.8660);
\draw[black,thin,dotted] (3/2, -0.8660)  -- (-1/2, 3*0.8660);
\draw[black,thin,dotted] (0, 0) --  (3/2, 3*0.8660);
\draw[black,thin] (0, 0) --  (-3/2, 0.8660);
\draw[black,thin,dotted] (5/2, 0.8660) --  (-3/2, 0.8660);
\draw[black,thin,dotted] (3/2, 3*0.8660) -- (-1/2, 3*0.8660);
\draw[black,thin,dotted] (-1/2, 3*0.8660) -- (-3/2, 0.8660);
\draw[black,thin] (3/2, 3*0.8660) -- (-3/2, 0.8660);
\draw[black,thin,dotted] (-3/2, 0.8660) --  (-1/2, -0.8660);
\draw[black,thin,dotted] (-3/2, 0.8660) -- (-5/2, -0.8660);
\draw[black,thin] (-3/2, 0.8660) -- (-3/2, -0.8660);
\draw[blue,thin] (-7/2, 0.8660) -- (-5/2, -0.8660) -- (-1/2, -0.8660) -- (1/2, 0.8660) -- (-1/2, 3*0.8660) -- (-5/2, 3*0.8660) -- cycle;
\draw[blue,thin] (3-7/2, 0.8660-1.7320) -- (3-5/2, -0.8660-1.7320) -- (3-1/2, -0.8660-1.7320) -- (3+1/2, 0.8660-1.7320) -- (3-1/2, 3*0.8660-1.7320) -- (3-5/2, 3*0.8660-1.7320) -- cycle;
\draw[blue,thin] (3-7/2, 0.8660+1.7320) -- (3-5/2, -0.8660+1.7320) -- (3-1/2, -0.8660+1.7320) -- (3+1/2, 0.8660+1.7320) -- (3-1/2, 3*0.8660+1.7320) -- (3-5/2, 3*0.8660+1.7320) -- cycle;
\draw[red,line width = 0.5mm] (0, 0) -- (3/2, -0.8660) -- (3, 0) -- (5/2, 0.8660) -- (3/2, 0.8660) -- (3/2, 3*0.8660) -- (0, 4*0.8660) -- (-1/2, 3*0.8660) -- (-3/2, 3*0.8660) -- (-3/2, 0.8660) -- (-3, 0) -- (-5/2, -0.8660) -- (-1/2, -0.8660) --cycle;
\draw[red,line width = 0.5mm] (0, 0) -- (3/2, -0.8660) -- (3, 0) -- (5/2, 0.8660) -- (3/2, 0.8660) -- (3/2, 3*0.8660) -- (0, 4*0.8660) -- (-1/2, 3*0.8660) -- (-3/2, 3*0.8660) -- (-3/2, 0.8660) -- (-3, 0) -- (-5/2, -0.8660) -- (-1/2, -0.8660) --cycle;
\end{tikzpicture}
\caption{The {\em hat} tile as a subset of the regular hexagonal tiling; if $a$ denotes the size of the regular hexagon edge, the tile area is $2\sqrt{3}a^2$ and the lengths of each edge is either $a,$ $\sqrt{3}a$ or $2a.$ Notice that all hat tiles edges are either edges (or half-edges) or heigths of a hexagonal lattice.}  
\label{TheHat}
\end{figure}
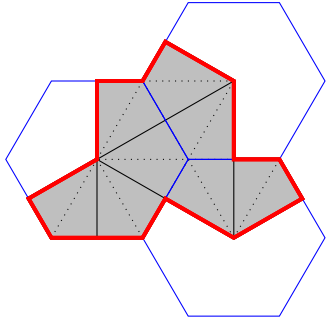

From a symmetry perspective, the hat tiling is a special case: as it is built only upon the edges and heights of the hexagonal lattice, all edge orientations of the tiling fall into the set of multiples of $\pi/6$ (see Fig. \ref{TheHat}). Despite this, at larger scales, the meta-tiles of the hat tiling are self-similar only after a rotation by an angle $\phi$ such that $\pi/\phi$ is irrational, and this is a key ingredient of the aperiodicity proof. Another peculiarity of the hat tilling is its low coordination number ($\simeq 2.3$). This means that 2/3 of its vertices are only connected to two different edges, so that in order to avoid zero-modes one has to include interactions that go beyond near neighbors.   

The paper is organized as follows: the second section presents the minimal geometric elements of the aperiodic tiling, described in detail in \cite{Smith-Aperiodic-2023, Smith-Chiral-2023}. While aperiodicity is granted by construction, hyperuniformity is here the consequence of a single-tile construction. The third section presents the minimal mechanical requirements that ensure both the rigidity and the integrity of a finite part of the plane tiling, and we study the behavior of a Hooke law approximations along a sequence of different length-scales. The simplest mechanical model endows the planar tiling with (i) the quadratic approximations of the pair-interactions (of NN type) along the edges of all polygons and the Kirkwood-Keating approximation of the angular interactions at all polygons vertices. We formulate a generic problem that allows the computation of the macroscopic stress by solving three Dirichlet boundary-value problems with homogeneous boundary conditions at a fixed macroscopic strain \cite{E-2007}. 

In order to account for the proximity of the macroscopic constitutive relation (Hooke tensor) to the subspace of isotropic fourth-order tensors we introduce an {\em isotropy index}. It measures the distance between a computed (anisotropic) Hooke tensor and the hyperplane of isotropic Hooke tensors and is defined as the Euclidean distance between the computed Hooke tensor and its orthogonal projection onto the hyperplane of isotropic tensors of Hooke type\footnote{Here, by Hooke type, we mean fourth-order tensors that respect both the two minor and the major symmetries.}, normalized by the norm of the computed Hooke tensor. The numerical results obtained using the discrete model are presented in the fourth section: we compute the isotropy index for several realizations along an increasing sequence of length scales and find that: (i) the isotropy index decreases along an increasing sequence of length scales, (ii) the numerical values of the set averages (at fixed length scales) of the Lamé constants evolve toward constant values and (iii) the dispersion of numerical results corresponding to various realizations at fixed length-scale decreases with respect to the length-scale. All these results point toward an isotropic macroscopic law, as expected. 

In view of a potential experiment, the fifth section concerns a second mechanical model that identifies the edges of the planar tiling with a one-dimensional continuum and accounts, along all polygon edges, for both compression-extension, shear and bending elastic energies in the simplest Timoshenko approximation — a classical setting for piecewise straight beam structures. The numerical results obtained in this case also show that the macroscopic elastic behavior tends toward an isotropic continuum Hooke law. Inclusion of the shear effect provides here a non-Cauchy part of the macroscopic Hooke tensor. However, while in realistic situations most of the elastic energy is stored in the bending term, the macroscopic stress contribution accounts in an explicit manner only for extension/compression and shear. For this reason, only in the case when the internal length-scale of the mechanical interactions (here $\sqrt{J/S}$) is of the same order of magnitude as the length-scale of the tiling we notice a behaviour compatible with convergence to an isotropic continuum. The practical situation when $\sqrt{J/S}$ is much lower than the length-scale of the single tile edges will be the subject of future work.      

\section{Geometry of the hat aperiodic tiling}
\label{Geometry}

The geometry of the aperiodic hat tiling is described in detail in \cite{Smith-Aperiodic-2023}: it is built by using a unique 13 edges polygonal tile (see Figure 1) and inflation rules that consist of refinement and substitutions using meta-tiles. For the technical details involved in the generation of larger aperiodic tilings, we refer the reader to \cite{Smith-Aperiodic-2023, Smith-Chiral-2023, Socolar-2023}. In the present work we used the results of \cite{Klee-2023} that allow to both create and handle efficiently very large (up to $10^6$ polygons and corresponding to $10^7$ vertices) aperiodic domains.

\subsection{Domains and boundary definitions}

\begin{figure}[!ht]
\centering
\begin{tikzpicture}[scale = 0.1]
\input{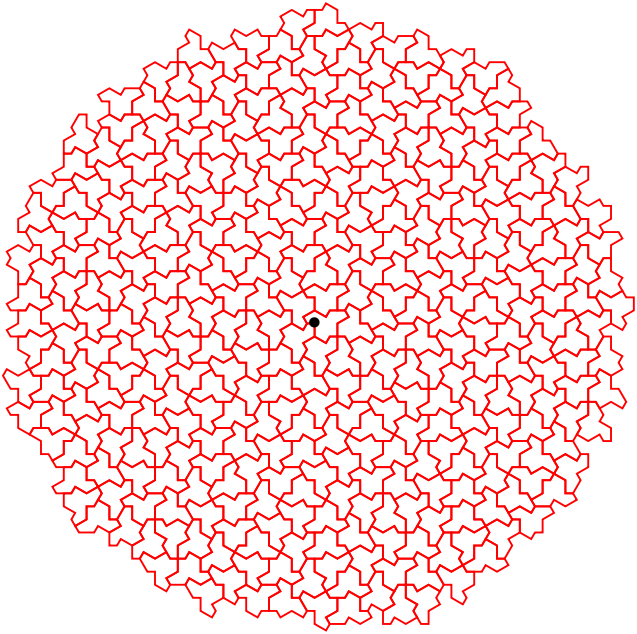}
\draw (0,0) node {$\bullet$}; 
\end{tikzpicture}
\caption{A typical small-size (363 polygons, 2231 vertices) circular subset of a aperiodic tiling. The black dot indicates the origin of coordinates in the plane.}  
\end{figure}

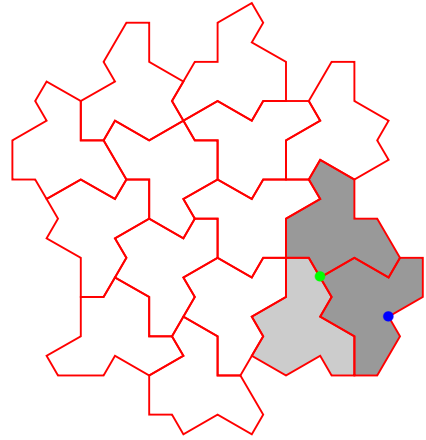
\begin{figure}[h!]
\centering
\begin{tikzpicture}[scale = 0.3]
\input{Figure3}
%\draw (0,0) node {$\bullet$}; 
\draw[green] (4.5,-2.59) node {$\bullet$}; 
\draw[blue]  (7.5,-4.35) node {$\bullet$}; 
\end{tikzpicture}
\caption{The green dot is a vertex for the two dark gray polygons but is not a vertex for the light gray polygon. In this case, we add an additional {\sl node/vertex} to the light gray polygon in order to avoid fracture/interpenetration.} 
\label{GeometricDetails}
\end{figure}

All computations presented here were performed on domains (realizations) obtained as the union of polygons for which the centroid lies in a given disk, whose center has various positions in the plane tiling. As an illustration, the domain in Figure 2 contains only the tiles for which the distance between the polygons centroid (i.e., the barycenter) and the origin of the coordinate system (the black dot) is less or equal to a fixed value, here $10a.$ Obviously, the hat-tile area is $2\sqrt{3}a^2$ and it is interesting to notice that, even if the tiling is aperiodic, all vertices orientations are multiples of $\pi/6.$ Despite this, the tiling is aperiodic \cite{Smith-Aperiodic-2023}.     

The next step is to distinguish between points located in the interior of a domain (as that illustrated in Figure 2) and those along its boundary. This highlights a first geometric peculiarity related to the hat tiling: as shown in Fig. 3, the tiling may contain vertices (green dot in Fig. 3) common to two adjacent polygons (the two dark gray polygons in Fig. 3) that are not vertices of the third one (the light gray polygon in Fig. 3) but only midpoints along one of its edges. This can occur either in the interior or along the boundary of the domain and may lead to fracture and/or interpenetration of the structure. To address this, we modify the connectivities in the structure and add an additional midpoint along the edges where this occurs. As a consequence some polygons of the (modified) planar tiling contain 14 edges, two of them of identical orientation. 

\begin{figure}[ht!]
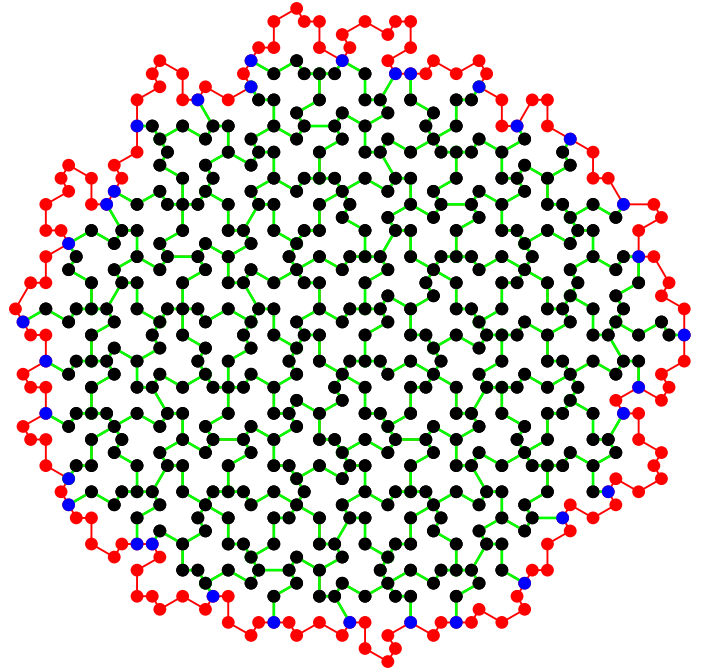

\centering
\begin{tikzpicture}[scale = 0.2]
\input{Figure31}
\input{Figure32}
\end{tikzpicture}
\caption{Interior vertices (black dots) and only boundary vertices connected to interior ones (blue dots) are accounted for in subsequent computations. Only connectivities represented in green account for the interaction between the bondary and interior vertices.}  
\end{figure}

As a second peculiarity, we notice that not all vertices located on the boundary are connected to vertices located in the interior of the domain. Thus, as a final step in defining the boundary of the domain, we select only those boundary vertices that are connected to interior vertices. Figure 4 illustrates the interior vertices (black dots), boundary vertices (blue dots), and connections involving at least one interior vertex (green lines) in a generic situation. Vertices not connected to interior ones (red dots), as well as the boundary edges connecting two boundary vertices (red lines) will not be taken into account in all subsequent computations. Other alternative choices of the domain and its boundary exist: as an example, one can define as interior points all vertices from all polygons and the boundary is obtained by adding all vertices that connect this polygonal domain to the rest of the tiling. We claim that, due to the equilibirum, these two alternative definitions provide the same results, in the limit of a large domains. 

\section{Discrete mechanical interactions and the macroscopic response}
\label{Mechanics}

\subsection{Pair and three-body interactions}
In order to investigate the elastic properties of the tiling structure we first endow the geometrical pattern with classical physical interactions. In the simplest setting, pair-interactions along the edges of the tiling will be considered. If $k_1$ denotes the stiffness of the NN-type pair-interaction between $P_i$ and $P_j$ - the end-points of the same edge with coordinates ${\bs X}^i$ and ${\bs X}^j$ respectively - and ${\bs r}^{ij}_0 = {\bs X}^j-{\bs X}^i,$ the elastic energy is defined as 
\begin{equation}                                                                                                                                                                                                                                            W_{ij}=\frac{k_1}{8|{\bs r}^{ij}_0|^2}\left[{\bs r}^{ij}\cdot{\bs r}^{ij}-{\bs r}^{ij}_0\cdot{\bs r}^{ij}_0\right]^2.                                                                                                                                                                                                                                                        \end{equation}
Hereabove, ${\bs r}^{ij}$ is defined in terms of the relative displacement ${\bs u}^j-{\bs u}^i$ through ${\bs r}^{ij}={\bs X}^j-{\bs X}^i+{\bs u}^j-{\bs u}^i= {\bs r}^{ij}_0+{\bs u}^j-{\bs u}^i$ so that, at second order with respect to displacement, we obtain the classical compression-extension elastic energy  
\begin{equation}                                                                                                                                                                                                                                            W_{ij}\simeq\frac{k_1}{2|{\bs r}^{ij}_0|^2}\left[({\bs u}^j-{\bs u}^i)\cdot{\bs r}^{ij}_0\right]^2=\frac{k_1}{2}\left[({\bs u}^j-{\bs u}^i)\cdot{\bs N}^{ij}\right]^2.
\label{Wij}
\end{equation}

It is well-known that by taking into account only pair-interactions on generic polygons one cannot insure the mechanical stability of the structure, due to zero-modes. The simplest ingredient that overcome this drawback is the addition of angular (or three-body) interactions between edges sharing a common vertex. If $(P_i,P_j)$ and $(P_i,P_k)$ are vertices of the same polygon the three-body elastic energy of the angular interaction at $P_i$ is
\begin{equation}
W_{ijk}=\frac{k_2}{2|{\bs r}^{ij}_0||{\bs r}^{ik}_0|}\left[{\bs r}^{ij}\cdot{\bs r}^{ik}-{\bs r}^{ij}_0\cdot{\bs r}^{ik}_0\right]^2
\end{equation}
which, after developement leads to the quadratic approximation
\begin{equation}
W_{ijk} 
\simeq  \frac{k_2}{2|{\bs r}^{ij}_0||{\bs r}^{ik}_0|}\left[({\bs u}^j-{\bs u}^i)\cdot{\bs r}^{ik}_0+({\bs u}^k-{\bs u}^i)\cdot{\bs r}^{ij}_0\right]^2.
\label{Wijk}
\end{equation}
This is known (see \cite{Starzewski-2002}) as the Kirkwood-Keating approximation of the three-body (angular) interaction. Notice that, if in the reference $P_i$ is the mid-point of the segment $(P_j,P_k)$ than $|{\bs r}^{ik}_0|=|{\bs r}^{ij}_0|=\frac{1}{2}|{\bs r}^{jk}_0|$ and $$
\left[({\bs u}^j-{\bs u}^i)\cdot{\bs r}^{ik}_0+({\bs u}^k-{\bs u}^i)\cdot{\bs r}^{ij}_0\right]^2=\frac{1}{4}\left[({\bs u}^k-{\bs u}^j)\cdot{\bs r}^{jk}_0\right]^2$$
so that 
(\ref{Wijk}) becomes $W_{jk}$ defined in (\ref{Wij}). 
 
If $N$ is the number of all vertices in the system, the total elastic energy is then 
\begin{equation}
W(\{{\bs u}^j\})= \sum_{\{i,j\}}W_{i,j}(\{{\bs u}^j\})+\sum_{\{i,j,k\}}W_{ijk}(\{{\bs u}^j\})
\label{W_Tot}
\end{equation}
where, $\{{\bs u}^j\}\in{\mathbb R}^{2N}$ (as ${\bs u}^j\in{\mathbb R}^2$) denotes the collection of all particles displacements, the first sum runs over all pair interactions and the second one over all angular interactions.

\subsection{Boundary-value problem and macroscopic elasticity}

For $n\in{\mathbb N},$ let $\Omega_n$ denote the domain formed by the union of all polygons with centroid in the disk with center at $O$ and radius $n.$ Assume that all edges and angles of all polygons in $\Omega_n$ are endowed with pair and three-body (angular) interactions and let $W_n(\{{\bs u}^j\})$ be the total elastic energy defined in (\ref{W_Tot}). Then, in the class of displacements that satisfy a homogeneous boundary condition, i.e.,
\begin{equation}
{\bs u}^i = {\bs E}{\bs X}^i,    
\label{HBC}
\end{equation}
at all vertices located on the boundary of $\Omega_n,$ 
there is an unique solution $\{{\bs u}^j\}$ of the minimization problem  
\begin{equation}
\{{\bs u}^j\} = \textrm{argmin} W_n(\{{\bs u}^j\}).
\label{MinProblem}
\end{equation}
Clearly, this solution depends on both $n$ and $O$ and is linear with respect to ${\bs E}.$ Thus, the minimum of the total elastic energy becomes a quadratic form in ${\bs E}$ and, by using the classical Hill lemma, one defines the macroscopic stress\footnote{Here the stress in defined per unit thickness so that the units for ${\bs S}$ are in $N/m.$} as 
\begin{equation}
{\bs S}_{O} = 
\lim_{n\rightarrow\infty}{\bs S}^{(n)}_{O}=
\lim_{n\rightarrow\infty}\left[\frac{1}{\textrm{area}(\Omega_n)}\frac{\partial W_n}{\partial {\bs E}}\right]=
(\lim_{n\rightarrow\infty}{\mathbb C}^{(n)}_{O})[{\bs E}].
\end{equation}
If this limit exists and it is independent on the realization (position of the center of $\Omega_n$ with respect to the tilling), it defines unambiguously the macroscopic elastic behavior of the hat tiling.  

The standard procedure to solve the minimization problem (\ref{MinProblem}) is to introduce {\em the fluctuation} vector defined as $\hat{\bs u}^j={\bs u}^j-{\bs E}{\bs X}^j$ so that the total energy becomes a quadratic form in $\hat{\bs u}$ and ${\bs E}{\bs X}$ defined as 
\begin{equation}                                                                                                                                                                                                                                                          
W_n(\{{\bs u}^j\})=\hat{W}_n(\{{\bs E{\bs X}^j}\},\{\hat{\bs u}^j\}).                                                                                                                                                                                                                                              \end{equation}

At a given length-scale ($n$ fixed) we obtain 
\begin{equation}
\label{macroS}
\begin{split}
 & {\bs S}_n = {\mathbb C}_n[{\bs E}] = \\ 
& = \frac{1}{\textrm{area}(\Omega_n)}
\left[
\frac{\partial \hat{W}_n    }{\partial \hat{\bs u}^j}
\frac{\partial \hat{\bs u}^j}{\partial {\bs E}  }+
\frac{\partial \hat{W_n}    }{\partial {\bs E}  }
\right]
=\frac{1}{\textrm{area}(\Omega_n)}
 \frac{\partial\hat{W_n}}{\partial{\bs E}} = \\
& =\frac{1}{\textrm{area}(\Omega_n)}
\left[
\sum_{\{i,j\}}    \frac{\partial \hat{W}_{ij}}{\partial{\bs E}}+
\sum_{\{i,j,k\}}\frac{\partial W_{ijk}}{\partial{\bs E}}
\right],
\end{split}
\end{equation}
where the equality on the second line above is the consequence of the fact that 
$\partial \hat{W}_n/\partial \hat{\bs u}^j$ vanish at equilibrium. From (\ref{Wij}) and (\ref{Wijk}) we obtain the explicit contributions of both pair-interactions and three-body interactions respectively, as 
\begin{equation}
 \frac{\partial \hat{W}_{ij}}{\partial {\bs E}}=
 k_1\left[({\bs E}{\bs r}^{ij}_0)\cdot{\bs r}^{ij}_0+(\hat{\bs u}^j-\hat{\bs u}^i)\cdot{\bs r}^{ij}_0\right]
 {\bs N}^{ij}\otimes{\bs N}^{ij}
 \label{k1_Contribution}
\end{equation}
and
\begin{align}
\begin{split}
 & \frac{\partial\hat{W}_{ijk}}{\partial {\bs E}} = 
 k_2
 \left[
 2({\bs E}{\bs r}^{ij}_0)\cdot{\bs r}^{ik}_0+(\hat{\bs u}^j-\hat{\bs u}^i)\cdot{\bs r}^{ik}_0
 \right. + \\
& \left.+ (\hat{\bs u}^k-\hat{\bs u}^i)\cdot{\bs r}^{ij}_0
 \right]({\bs N}^{ij}\otimes{\bs N}^{ik}+
         {\bs N}^{ik}\otimes{\bs N}^{ij}).
\end{split}
\end{align}

Notice that, while the pair interaction term provide as expected only a Cauchy type\footnote{Here by Cauchy type we understand a fourth-order tensor which is symmetric with respect to all possible permutations of subscripts.} contribution to the macroscopic stress tensor \cite{Itin-2002,Itin-2023}, the angular interaction provide contributions to both the Cauchy and non-Cauchy parts. 

For practical purposes, as the Hooke law in two dimensions can be written as (Voigt notation) 
\begin{equation}
\begin{pmatrix} S_{11} \cr S_{22}\cr S_{12}  \end{pmatrix}
=
\begin{pmatrix}
 H_{11} & H_{12} & H_{16}\cr
 H_{12} & H_{22} & H_{26}\cr
 H_{16} & H_{26} & H_{66}
\end{pmatrix}
\begin{pmatrix} E_{11} \cr E_{22}\cr 2E_{12}  \end{pmatrix},
\label{Voigt}
\end{equation}
one has to solve three independent problems, corresponding to three homogeneous Dirichlet boundary conditions with a fixed macroscopic strain (Voigt notation) 
\begin{equation}\label{typeE}
 {\bs E}\in \left\{ 
\begin{pmatrix} 1 \cr 0\cr 0  \end{pmatrix},  
\begin{pmatrix} 0 \cr 1\cr 0  \end{pmatrix},
\begin{pmatrix} 0 \cr 0\cr 1/2\end{pmatrix}
\right\}.
\end{equation}
For ${\bs E}$ fixed if $\hat{\bs u}({\bs E})$ is the solution of the minimization problem, using  (\ref{macroS}) we obtain from (\ref{Voigt}) a column of the computed approximation of the Hooke tensor. 

\subsection{Anisotropy index}

In two dimensional elasticity\footnote{For convenience, here we do not regard the two-dimensional elasticity as the plane-stress or plane-strain approximation of the three dimensional elasticity; however, with minor modifications, one can extend the interpretation of the two-dimensional constitutive relation obtained as plane-stress/plane-strain approximations.}, the set of isotropic fourth-order tensors of Hooke type form a two-dimensional vector space (hyperplane) spanned by linear combinations of (Voigt notation) the two orthogonal tensors
\begin{equation}
 {\mathbb H}_1 = \frac{1}{2}
 \begin{pmatrix}
 1 & 1 & 0\cr
 1 & 1 & 0\cr
 0 & 0 & 0
 \end{pmatrix} \quad \textrm{and}\quad
 {\mathbb H}_2 = \frac{1}{2\sqrt{2}}
 \begin{pmatrix*}[r]
 1 & -1 & 0\cr
-1 & 1 & 0\cr
 0 & 0 & 1
 \end{pmatrix*}.
\end{equation}
Hereabove, the orthogonality is defined by using the scalar product of fourth-order tensors of Hooke type induced by that of linear applications from $M_2({\mathbb R})$ to itself (fourth-order tensors with minor and major symmetries), i.e. 
${\mathbb A}::{\mathbb B} = A_{ijkl} B_{ijkl}$ so that, by using Voigt notation and symmetries described by subscripts permutations, we obtain
\begin{equation}
\begin{split}
 \|{\mathbb H}\|^2 &= \langle{\mathbb H}::{\mathbb H}\rangle =\\
 &= H_{11}^2+H_{22}^2+2H_{12}^2+4(H_{16}^2+H_{26}^2+H_{66}^2).
\end{split}
\label{HookeNorm}
\end{equation}
It is easy to check that
\begin{equation}
\begin{pmatrix*}[c]
\lambda + 2\mu & \lambda & 0\cr
\lambda & \lambda + 2\mu & 0\cr
0 & 0 & \mu
\end{pmatrix*}
= 2[(\lambda+\mu){\mathbb H}_1 + \sqrt{2}\mu{\mathbb H}_2]. 
\end{equation}
Since any fourth-order tensor of Hooke type ${\mathbb H}$ can be written as a sum of three terms ${\mathbb H}=\alpha_1 {\mathbb H}_1 + \alpha_2 {\mathbb H}_2 + {\mathbb H}^\perp$ such that ${\mathbb H}^\perp ::{\mathbb H}_1=0$ and ${\mathbb H}^\perp ::{\mathbb H}_2=0,$ the normalized scalar quantity 
\begin{align}
\begin{split}
 I({\mathbb H})=\frac{\|{\mathbb H}^\perp\|}{\|{\mathbb H}\|} & = \frac{1}{{\|{\mathbb H}\|}}\left[4(H_{16}^2+H_{26}^2)+\frac{1}{2}(H_{11}-H_{22})^2+\right. \\
   & \left. + \frac{1}{8}(H_{11}+H_{22}-2H_{12}-4H_{66})^2\right]^{1/2},
\end{split}
\label{I(H)}
\end{align}
further called {\sl anisotropy index}, measure the (normalized) distance between the Hooke tensor ${\mathbb H}$ and the vector subspace of isotropic Hooke tensors. By construction, we obviously have $I({\mathbb H})=0$ if and only if ${\mathbb H}$ is isotropic. 

\section{Implementation and numerical results}

We have implemented the above described procedure along a sequence of length scales on circular domains $\Omega_n$ with diameter $an$ in the range $(10a,300a)$ and, at each fixed length-scale (i.e., $n$ fixed) on 10 different realizations that differ by the position of the center of the circular domain with respect to the planar tiling. As previously mentioned, the domains are selected from a very large tiling previously generated using the method in \cite{Klee-2023} by collecting all the polygons with centroid in the disk centered at various positions. 

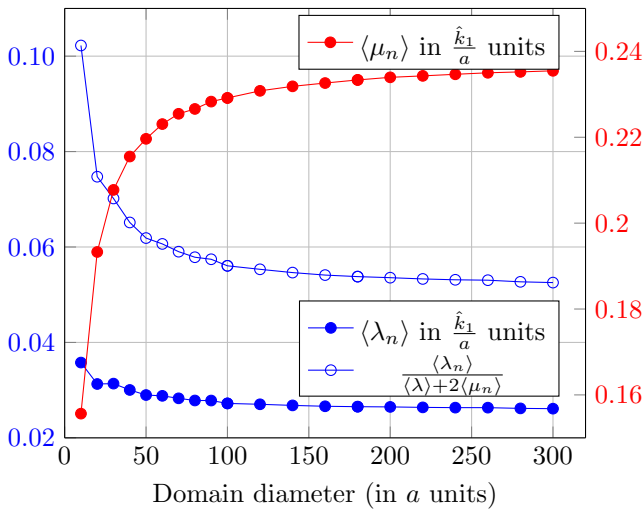
\begin{figure}[h!]
\centering
\begin{tikzpicture}[scale=1.0]
\begin{axis}
[xmin = 0, xmax = 320, ymin = 0.02, ymax = 0.11, 
axis y line*=left, xlabel={$n$}, xlabel near ticks,
y tick label style={/pgf/number format/zerofill,color=blue}, scaled y ticks=false,
ytick={0.02, 0.04, 0.06, 0.08, 0.10},grid=both,
legend style={at={(0.95,0.32)},anchor = north east},
xlabel={Domain diameter (in $a$ units)}
]
\addplot[color=blue,mark=*] coordinates{
( 10.000000 , 0.035778 )
( 20.000000 , 0.031287 )
( 30.000000 , 0.031366 )
( 40.000000 , 0.030038 )
( 50.000000 , 0.028972 )
( 60.000000 , 0.028805 )
( 70.000000 , 0.028283 )
( 80.000000 , 0.027829 )
( 90.000000 , 0.027821 )
( 100.000000 , 0.027216 )
( 100.000000 , 0.027216 )
( 120.000000 , 0.027034 )
( 140.000000 , 0.026800 )
( 160.000000 , 0.026617 )
( 180.000000 , 0.026532 )
( 180.000000 , 0.026532 )
( 200.000000 , 0.026484 )
( 220.000000 , 0.026385 )
( 240.000000 , 0.026332 )
( 260.000000 , 0.026328 )
( 280.000000 , 0.026173 )
( 300.000000 , 0.026116 )
};
\addlegendentry{$\langle\lambda_n\rangle$ in $\frac{\hat{k}_1}{a}$ units}
\addplot[color=blue,mark=o] coordinates{
( 10.000000 , 0.102234 )
( 20.000000 , 0.074732 )
( 30.000000 , 0.070173 )
( 40.000000 , 0.065142 )
( 50.000000 , 0.061872 )
( 60.000000 , 0.060644 )
( 70.000000 , 0.059020 )
( 80.000000 , 0.057855 )
( 90.000000 , 0.057429 )
( 100.000000 , 0.056056 )
( 100.000000 , 0.056056 )
( 120.000000 , 0.055323 )
( 140.000000 , 0.054637 )
( 160.000000 , 0.054115 )
( 180.000000 , 0.053793 )
( 180.000000 , 0.053793 )
( 200.000000 , 0.053569 )
( 220.000000 , 0.053306 )
( 240.000000 , 0.053127 )
( 260.000000 , 0.053043 )
( 280.000000 , 0.052702 )
( 300.000000 , 0.052541 )
};
\addlegendentry{$\frac{\langle\lambda_n\rangle}{\langle\lambda\rangle + 2\langle\mu_n\rangle}$}
\end{axis}
\begin{axis}[
xmin = 0, xmax = 320,ymin = 0.15, ymax = 0.25,
hide x axis, axis y line*=right,
y tick label style={/pgf/number format/precision=4,color=red},
legend style={at={(0.95,0.985)},anchor = north east}
]
\addplot[color=red,mark=*] coordinates{
( 10.000000 , 0.155635 )
( 20.000000 , 0.193287 )
( 30.000000 , 0.207740 )
( 40.000000 , 0.215509 )
( 50.000000 , 0.219632 )
( 60.000000 , 0.223067 )
( 70.000000 , 0.225463 )
( 80.000000 , 0.226592 )
( 90.000000 , 0.228301 )
( 100.000000 , 0.229151 )
( 100.000000 , 0.229151 )
( 120.000000 , 0.230808 )
( 140.000000 , 0.231855 )
( 160.000000 , 0.232622 )
( 180.000000 , 0.233346 )
( 180.000000 , 0.233346 )
( 200.000000 , 0.233953 )
( 220.000000 , 0.234292 )
( 240.000000 , 0.234660 )
( 260.000000 , 0.235009 )
( 280.000000 , 0.235220 )
( 300.000000 , 0.235470 )
};
\addlegendentry{$\langle\mu_n\rangle$  in $\frac{\hat{k}_1}{a}$ units}
\end{axis}
\end{tikzpicture}$\qquad$
\caption{Evolution of the (set) average values for the Lamé constants $\langle\lambda_n \rangle,$ $\langle\mu_n\rangle$ (in $\frac{\hat{k}_1}{a}$ units) and the Poisson coefficient $\nu_n =\frac{\langle\lambda_n +2 \mu_n\rangle}{\langle\lambda_n\rangle}$ as a function of length-scale.} 
\label{LaMuNu}
\end{figure}

Since the hat tiling contains segments with three different lengths a physicaly realistic assumption is to use the pair-interaction stiffnesses $k_1$ in (\ref{Wij}) proportional to the inverse of the spring length, an assumption reminiscent from the one-dimensional continuum theory where a beam of length $L$ stiffness is $k = \frac{ES}{L}.$ The physical meaning of this assuption is that {\em the springs represent the same material}. Thus for each pair interaction the stiffness involved in $W_{ij}$ is written as $k_{ij}=\hat{k}_1/|r^{ij}_0|$ and $\hat{k}_1$ (fixed) defines now a force scale. We can assume, without loosing the generality, that $\hat{k}_1 = 1$ so that the normalized macroscopic response depends only on the non-dimensional $a k_2/\hat{k}_1$ ratio and, as already noted, becomes singular when $k_2\rightarrow 0.$ As a consequence the results presented in Figures \ref{LaMuNu} and \ref{Index} are expressed in $\hat{k}_1/a$ units. 

\input{IsoIndexNNAN}

At each realization of a given length scale ($n$ fixed) we compute the (approximate) Lamé constants $\lambda_n={\mathbb C}^{(n)}_{12}$ and $\mu_n={\mathbb C}^{(n)}_{66}$ and their set average $\langle\lambda\rangle$ and $\langle\mu\rangle.$ The obtained results are illustrated in Figure \ref{LaMuNu} where in addition we also represent the Poisson coefficient, computed using the two-dimensional elasticity definition $\nu_n = \frac{\langle\lambda_n\rangle}{\langle\lambda_n+2\mu_n\rangle}.$   

Figure \ref{Index} shows both the individual values of the anisotropy index as defined in (\ref{I(H)}) for all realizations (in red) at all length scales considered, as well as the set average at a fixed length scale (in black). The numerical results show that the evolution of the set average of both Lamé constants ${\lambda_n, \mu_n}$ and $I(\mathbb{H})$ is consistent with an isotropic behavior, as conjectured. Its set average decreases from $6.8 \times 10^{-2}$ to $4.3 \times 10^{-4}$ (more than two orders of magnitude) as the length scale increases from $10a$ to $300a.$

It is worth noting that, while the numerical results for both the Lamé constants and the anisotropy index suggest convergence towards an isotropic material, this is relatively slow with respect to the length scale. The primary reason for this behavior is associated with the complexity of the aperiodic construction — a method employing inflation. The inflation pattern is structured in such a way that, as a function of the inflation index, the sequence of successive structures contain a substantial number of polygons. For example, the first 7 inflation steps yield, respectively, ${4, 25, 169, 1156, 7921, 54289, 372100}$ polygons (see \cite{Klee-2023}). Consequently, the orientations of the self-similar meta-tiles change only 7 times within this sequence. Thus, to achieve an approximation where the isotropy index decreases to much lower values requires very large areas, i.e., significant length scales. From this perspective, our most extensive computation, performed on a disk with a diameter of $300a$, involves $2 \times 10^4$ polygons, corresponding to approximately $2.3 \times 10^5$ degrees of freedom (DOF). The computation of the macroscopic constants takes several hours (on a large cluster) and, despite being two-dimensional, is slow due to the non-local nature of the angular interaction, resulting in a complex pattern in the sparse rigidity matrix. Nevertheless, the isotropic nature of the macroscopic law becomes apparent as the anisotropy index decreases by two orders of magnitude along the increasing length scales considered.

The left column of Figure \ref{fig:exp2} illustrates (in blue) the homogeneous part of the displacement field ($\bs{u}=\bs{E}\bs{x}$) superimposed on the reference configuration (in red) for the three cases in (\ref{typeE}). 

\clearpage
\newgeometry{top=0.2in,left=1in,bottom=0.5in}
\afterpage{\restoregeometry}
\begin{strip}\centering 
        \begin{tabular}{|c | c|} 
        \hline
            \includegraphics[scale =0.06]{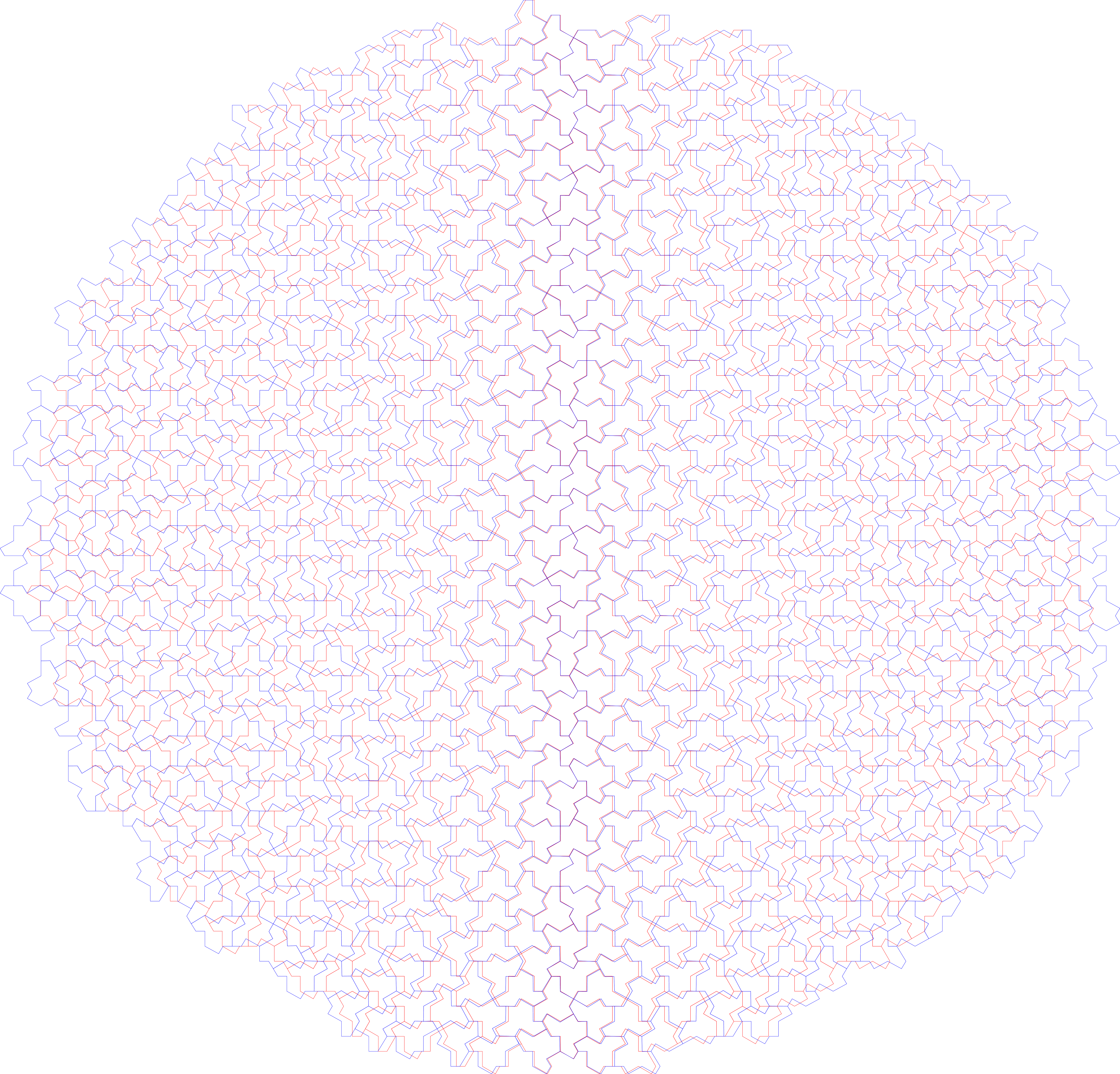}\qquad&
            \includegraphics[scale =1.6 ]{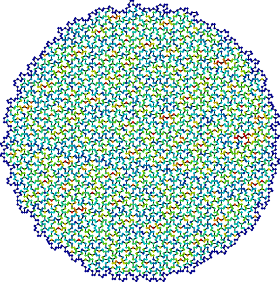}\cr\hline
            \includegraphics[scale =0.06]{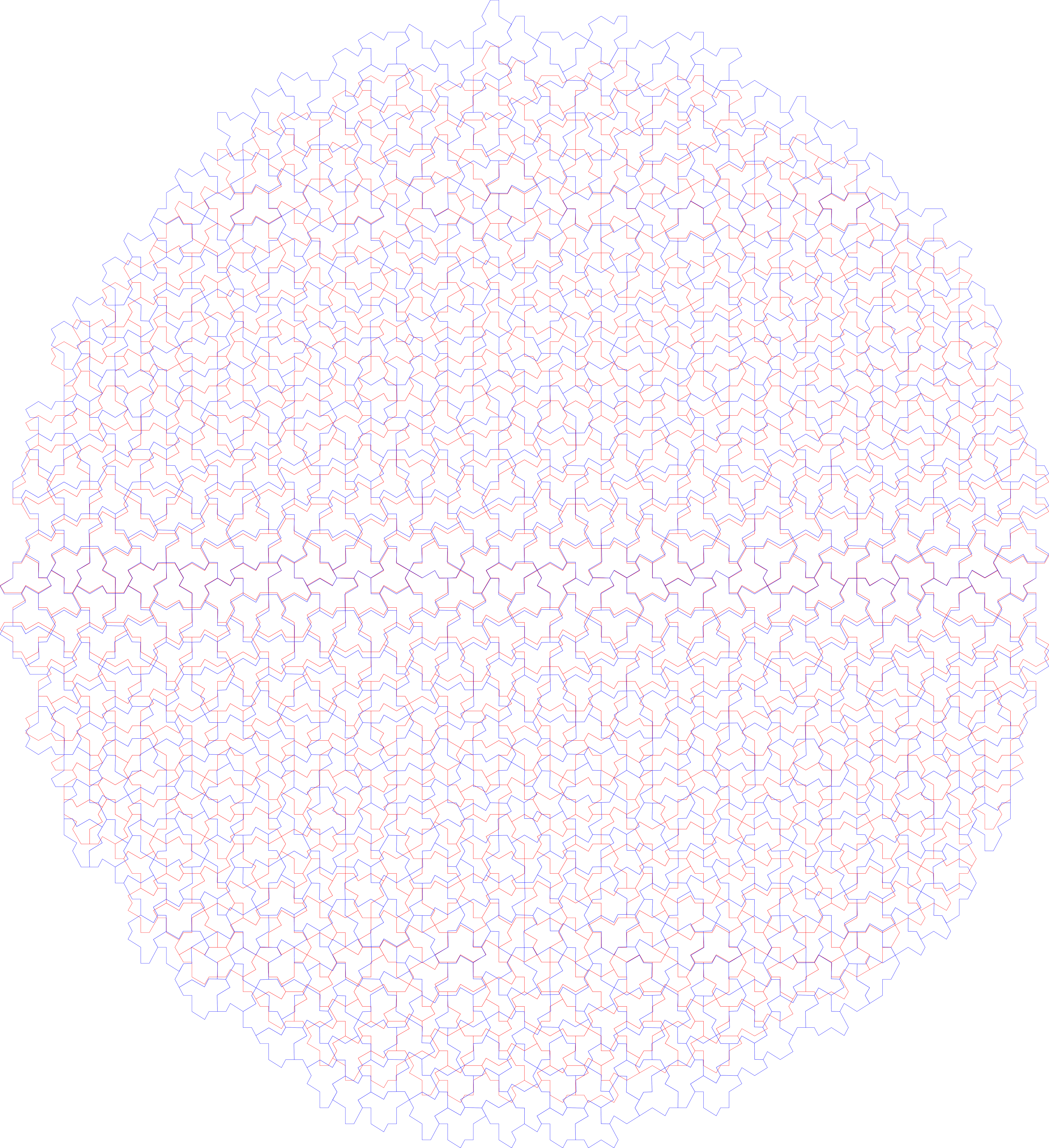}\qquad&
            \includegraphics[scale =1.6 ]{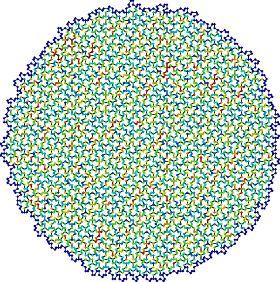}\cr\hline
            \includegraphics[scale =0.06]{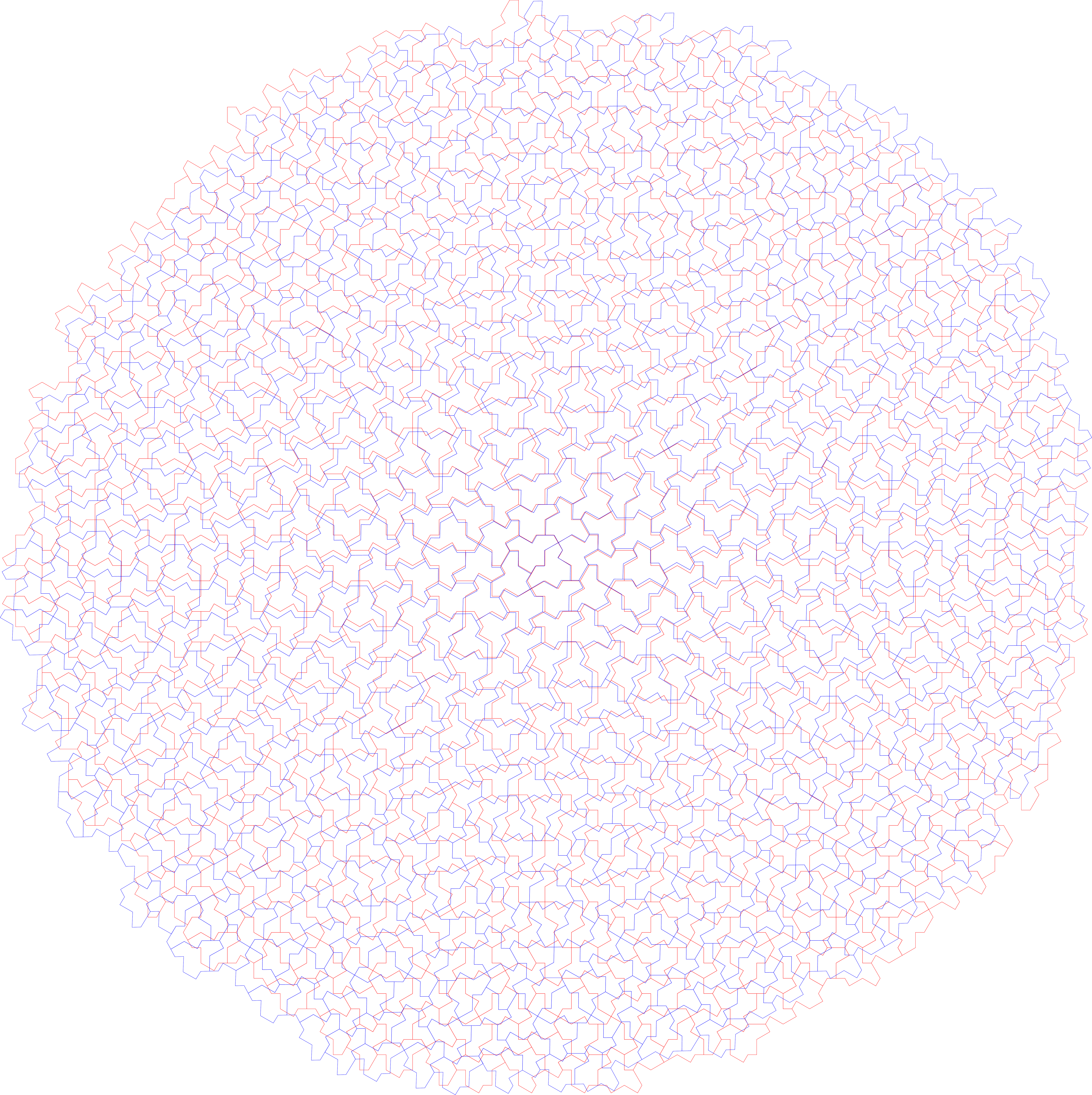}\qquad&
            \includegraphics[scale =1.6 ]{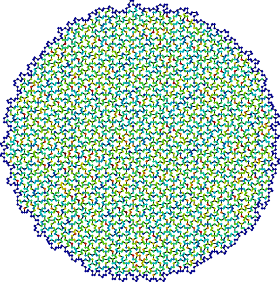}\cr
            \hline
        \end{tabular}
        \captionsetup{width=0.8\linewidth}
        \captionof{figure}{Illustration of a (small-scale) polygonal domain subset of the hat tiling; here, $r=30a$, an area which is one order order of magnitude lower than the largest realizations we have implemented. Left column: the reference (in red) and the homogeneous part (in blue) of the macroscopic deformation (see formula \ref{typeE}). Right column: the norm of the fluctuation vector represented by using a colormap from zero fluctuation in blue (as is the case along the boundary) to maximum values (in red) representing the highest deviations from the homogeneous part (high mobility points).}
        \label{fig:exp2}
\end{strip} 

\restoregeometry

The right column represents the norm of the fluctuation vector $|\hat{\bs{u}}|$ at each vertex of a small-scale domain with a radius of $30a$, using a normalized color scale where blue represents $|\hat{\bs{u}}|=0$ and red represents the maximum norm (indicating high mobility points).

\section{Toward an experiment: extension/compression, shear and bending elastic energies}

\subsection{The hat tiling as a one-dimensional continuum}

From the perspective of a future experiment, the assumption of pair-interactions can be regarded as a discrete approximation of the extension/compression property of a one-dimensional continuum, but the angular interaction does not have a simple continuum analog. To address the theoretical aspects of such an experiment, we will further investigate the same geometric structures but now endow them with a continuum structure at the microscale. By using additive fabrication, the geometric structure can be naturally obtained with dominant elastic properties from elastic extension/compression and bending. However (see (\ref{Ext_contribution})) this simplest setting will only provide a macroscopic Hooke tensor of Cauchy type \cite{Itin-2023}, i.e., with $\lambda=\mu$ so that\footnote{While in three-dimensional elasticity $\lambda=\mu$ implies $\nu=1/4,$ in two dimensions $\lambda=\mu$ leads to $\mu = 1/3$.} $\nu=1/3.$ In order to avoid this we adopt here the general framework of the Timoshenko model, thus including also the shear energy.  

To formalize this setting, let $s\in(0, L)$ be the arc length along an arbitrarily (but fixed) edge of the hat tiling, where $L$ denotes its length. If $(\bs{\tau}, \bs{t})$ defines the (fixed) local frame along the longitudinal and transverse directions of the fixed edge and $\bs{u}(s)$ denotes the displacement field along the barycenter line, then the longitudinal and transversal displacements are $\bs{u}\cdot\bs{\tau}$ and $\bs{u}\cdot\bs{t}$.

Within the framework of the Timoshenko theory the total elastic energy of an edge is the sum of the extension/compression, shear and bending elastic energies, i.e., 
\begin{equation}
\begin{split}
W_{\textrm{edge}}({\bs u},\omega)= &\frac{ES}{2}\!\!\int_{0}^{L}
\!\!\!\left(\frac{d{\bs u}}{ds}\cdot{\bs \tau}\right)^2\! ds + 
\frac{GS_2}{2}\!\!\int_{0}^{L}
\!\!\!\left(\frac{d{\bs u}}{ds}\cdot{\bs t}-\omega \right)^2\!ds +\\
+ & \frac{EJ}{2}\!\!\int_{0}^{L}
\!\!\!\left(\frac{d\omega}{ds}\right)^2\!ds,
\end{split}
\end{equation}
where $E$ denotes the Young modulus, $S$ the cross-section area $G=\frac{E}{2(1+\nu)},$ $S_2= \beta S$ (dependent on the shape of the cross-section) and $J$ the quadratic moment of the cross-section. As previously, the total elastic energy $W_n$ of a polygonal domain $\Omega_n$ is defined as the sum of individual elastic energies over all edges in the domain. In contrast to the previous model, including extension/compression and bending introduces an internal length scale $\sqrt{J/S}$ in the mechanical model at the micro-scale. 

\subsection{Boundary-value problem and interpolation}

The geometric definitions of the domain and its boundary are those of Section 2. As now the elastic energy incorporates a bending term, the homogeneous boundary condition in (\ref{HBC}):
$${\bs u}={\bs E}{\bs X},$$
has to be supplemented with additional information regarding either the rotation or bending moment (or a combination of both) at vertices located on the boundary. Two conditions have special significance: 
\begin{itemize}
\item[(i)] No rotation at boundary vertices, i.e., $\omega=0$ boundary condition; this condition is appropriate when the polygonal network is homogeneously deformed but its boundary is clamped in a larger frame, a situation close to experiment.  
\item[(ii)] Zero bending moment condition, i.e. (assuming identical $EJ$ for all edges which means same material in the tilling), $\omega'=0.$ This condition is appropriate when the rotation is free i.e., when ball (or pivot) joints are located at boundary vertices, a quite challenging experimental setting.
\end{itemize}

In the following, we shall focus on the first of these two conditions as it is easier to implement in an experiment.  

As the total elastic energy $W_n$ is quadratic with respect to the unknown fields $({\bs u},\omega)$ in the class of fields that satisfy the boundary conditions (\ref{HBC}) and $\omega=0$ (no rotation) at the boundary vertices of a polygonal domain $\Omega_n$ there is an unique field that realizes
\begin{equation}
({\bs u},\omega) = \textrm{argmin} W_n({\bs u},\omega).
\label{MinProblemRDM}
\end{equation}

As in the third section, we introduce the fluctuation vector field, which for each edge can be written as 
$\hat{\bs u}(s) = {\bs u}(s) - {\bs E}{\bs X}(s)$ and notice that it vanishes at vertices located on the boundary. Since $\frac{d{\bs X}}{ds}={\bs\tau}$ is constant along an edge, the elastic energy of an edge becomes
\begin{equation}
\begin{split}
& \hat{W}_{\textrm{edge}}({\bs E},\hat{\bs u},\omega)  = \frac{ES}{2}\int_0^{L}
\left[({\bs E}{\bs\tau}+\frac{\hat{\bs u}}{ds})\cdot{\bs\tau}\right]^2 ds + \\
& + \frac{GS_2}{2}\int_0^L\left[({\bs E}{\bs\tau}+\frac{\hat{\bs u}}{ds})\cdot{\bs t}-\omega\right]^2 ds + \frac{EJ}{2}\int_0^L(\frac{d\omega}{ds})^2 ds
\label{EdgeEnergy}
\end{split}
\end{equation}
and accounting for all edges over all polygons in the domain, the problem (\ref{MinProblemRDM}) becomes 
\begin{equation}
(\hat{\bs u},\omega) = \textrm{argmin} \hat{W}_n({\bs E}, \hat{\bs u},\omega).
\end{equation}

Once again, the unique solution of this problem $(\hat{\bs u}({\bs E}),\omega({\bs E}))$ depends linearily on ${\bs E}$ so that, the minimum of the total elastic energy is a quadratic form with respect to ${\bs E.}$ 

Further, the macroscopic stress is obtained as the partial derivative of the minimum elastic energy with respect to ${\bs E}$ and again, since the partial derivatives with respect to $\hat{\bs u}$ and $\omega$ vanish, for a given realization (at fixed legth-scale) we obtain
\begin{equation}
{\bs S}_n=\frac{1}{{\textrm{area}}(\Omega_n)}\frac{\partial\hat{W}_n}{\partial{\bs E}}={\mathbb C}_{n}[{\bs E}]
\label{Macro_RDM}
\end{equation}
If $\lim_{n\rightarrow\infty}{\mathbb C}_n$ exists and is independent on the realization, we have obtained the macroscopic elastic law. 

Using (\ref{EdgeEnergy}), since ${\bs\tau}$ is constant along an edge, the partial derivative in (\ref{Macro_RDM}) contains two contributions  
\begin{itemize}
 \item Extension/compression contribution : 
\begin{equation}
\begin{split}
&ES\left[(L({\bs E}{\bs\tau})\cdot{\bs\tau}+\int_0^L(\hat{\bs u}'\cdot{\bs\tau}))ds \right]{\bs\tau}\otimes{\bs\tau}=\\
& = 
ES\left[({\bs E}({\bs X}^j-{\bs X}^i)+\hat{\bs u}^j-\hat{\bs u}^i)\cdot{\bs\tau}\right]{\bs\tau}\otimes{\bs\tau}
\end{split}
\label{Ext_contribution}
\end{equation}
where we used ${\bs X}^j$ and ${\bs X}^i$ for the positions vectors of the end-points of the fixed edge (with respect to an arbitrary origin) and $\hat{\bs u}^j$ and $\hat{\bs u}^i$ are their respective fluctuations at equilibirum. This is the same expression in (\ref{k1_Contribution}). 
\item Shear contribution
\begin{equation}
\frac{GS_2}{2}\left[({\bs E}({\bs X}^j-{\bs X}^i)+\hat{\bs u}^j-\hat{\bs u}^i)\cdot{\bs t}\right]\left({\bs\tau}\otimes{\bs t}+{\bs t}\otimes{\bs\tau}\right)
\label{Shear_contribution}
\end{equation}
\end{itemize}

We underline here that, as previously mentioned, (\ref{Ext_contribution}) contibutes only to the Cauchy part of the macroscopic stress while the shear term (\ref{Shear_contribution}) contributes to both the Cauchy and non-Cauchy parts of it. Moreover, while the bending energy contributes to minimization, its contribution to the macroscopic stress remains only implicit. 

\subsection{Numerical results}

To minimize the total energy with respect to $(\hat{\mathbf{u}},\omega)$ one has to solve a (very large) set of coupled ordinary differential equations (ODE). We have used here the simplest interpolation scheme, i.e. linear interpolation for both ${\bs u}\cdot{\bs\tau}$ and $\omega,$ but quadratic for ${\bs u}\cdot{\bs t}.$ Notice that this is not a drastic simplification: for instance, as along each edge there is no external effort, the equilibirum equations for the extension/compression force provide constant normal effort and thus, linear longitudinal displacement. The numerical computations were conducted on the same realizations as those in Section 3, but now using the material length-scale factor $\sqrt{J/S} = a/2$ and $ES/GS_2=1/2.$ While the non-dimensional ratio $ES/GS_2$ is realistic, decreasing the material length-scale to much lower values induces an increasing contribution of the bending energy which, as previously noticed, is accounted for only in an implicit form in (\ref{Ext_contribution}) and (\ref{Shear_contribution}). This situation is an analog of the one-dimensional case where a horizontal zig-zag chain is submitted to a longitudinal force: its stiffness decrease when the bending energy dominates the extension/compression energy.

\begin{figure}[h!]
\centering
\begin{tikzpicture}[scale=1]
\begin{axis}
[xmin = 0, xmax = 320, ymin = 0.04, ymax = 0.11, 
axis y line*=left, xlabel={$n$}, xlabel near ticks,
y tick label style={/pgf/number format/zerofill,color=blue}, scaled y ticks=false,
ytick={0.04, 0.05, 0.06, 0.07, 0.08, 0.09, 0.1, 0.11},
legend style={at={(0.395,0.95)},anchor = north west},grid=both,
xlabel={Domain radius (in $a$ units)}
]
\addplot[color=blue,mark=*,mark size = 2] coordinates
{
( 10.000000 , 0.113636 )
( 20.000000 , 0.092176 )
( 30.000000 , 0.084142 )
( 40.000000 , 0.078963 )
( 50.000000 , 0.076018 )
( 60.000000 , 0.074838 )
( 70.000000 , 0.073165 )
( 80.000000 , 0.072544 )
( 90.000000 , 0.071486 )
( 100.000000 , 0.070634 )
( 120.000000 , 0.069750 )
( 140.000000 , 0.069285 )
( 160.000000 , 0.068767 )
( 180.000000 , 0.068385 )
( 180.000000 , 0.068385 )
( 200.000000 , 0.068061 )
( 220.000000 , 0.067914 )
( 240.000000 , 0.067682 )
( 260.000000 , 0.067492 )
( 280.000000 , 0.067309 )
( 300.000000 , 0.067121 ) 
 };
 \addlegendentry{$\frac{\langle\lambda\rangle}{\langle\lambda\rangle + 2\langle\mu\rangle}$}
 \addplot[color=blue,mark=o] coordinates
 {
( 10.000000 , 0.044937 )
( 20.000000 , 0.051487 )
( 30.000000 , 0.053822 )
( 40.000000 , 0.055043 )
( 50.000000 , 0.055775 )
( 60.000000 , 0.056316 )
( 70.000000 , 0.056657 )
( 80.000000 , 0.056889 )
( 90.000000 , 0.057142 )
( 100.000000 , 0.057268 )
( 120.000000 , 0.057530 )
( 140.000000 , 0.057715 )
( 160.000000 , 0.057843 )
( 180.000000 , 0.057948 )
( 200.000000 , 0.058043 )
( 220.000000 , 0.058119 )
( 240.000000 , 0.058174 )
( 260.000000 , 0.058231 )
( 280.000000 , 0.058262 )
( 300.000000 , 0.058296 )
};
\addlegendentry{$\langle\mu_n\rangle$ ($ES/a$ units)}
\end{axis}
\begin{axis}[
xmin = 0, xmax = 320,ymin = 0.008, ymax = 0.012,
hide x axis, axis y line*=right,
y tick label style={/pgf/number format/precision=3,color=red}, scaled y ticks=false,
legend style={at={(0.95,0.75)},anchor = north east},
ytick={0.008, 0.009, 0.010, 0.011, 0.012}]

\addplot[color=red,mark=*] coordinates{
( 10.000000 , 0.011611 )
( 20.000000 , 0.010467 )
( 30.000000 , 0.009891 )
( 40.000000 , 0.009438 )
( 50.000000 , 0.009178 )
( 60.000000 , 0.009111 )
( 70.000000 , 0.008945 )
( 80.000000 , 0.008900 )
( 90.000000 , 0.008799 )
( 100.000000 , 0.008705 )
( 120.000000 , 0.008627 )
( 140.000000 , 0.008593 )
( 160.000000 , 0.008543 )
( 180.000000 , 0.008507 )
( 200.000000 , 0.008478 )
( 220.000000 , 0.008470 )
( 240.000000 , 0.008446 )
( 260.000000 , 0.008429 )
( 280.000000 , 0.008409 )
( 300.000000 , 0.008389 )
};
\addlegendentry{$\langle\lambda_n\rangle$ ($ES/a$ units)}
\end{axis}
\end{tikzpicture}$\qquad$
\caption{Evolution of the set-average Lamé constants $\langle\lambda_n \rangle,$  $\langle\mu_n\rangle$ (in $ES/a$ units) and the Poisson coefficient $\frac{\langle\lambda_n\rangle+\langle\mu_n\rangle}{\langle\lambda_n\rangle}$ as a function of the length-scale.}
\label{LaMuNuRDM}
\end{figure}

\input{IsoIndexExtBend}

The numerical results obtained are presented in Figures \ref{IndexRDM} and \ref{LaMuNuRDM}. It's noteworthy that the evolution of elastic constants ans the anisotropy index is similar to that observed in the discrete mechanical model, as both the (set average) distance to the hyperplane of the isotropic tensor and the dispersion are decreasing functions with respect to the length-scale. In the largest computation result, performed on a domain with a diameter of $300a$, the set-average distance to the hyperplane of isotropic tensors is achieved at $3\times 10^{-4}.$ Once again, this behavior is consistent with a macroscopic elastic constitutive relation of an isotropic material, as expected. 

\section{Conclusions and perspectives}

We have explored the macroscopic mechanical behavior of the aperiodic tiling of the plane that uses a single (the hat) tile. By construction, this is a hyperuniform material with a single length-scale and despite the fact that all edges of all polygons in the tiling are located along the symmetry lines of an hexagonal tiling of the plane, the tiling does not have any particular orientational symmetry. Under suitable assumptions, such as the uniformity of material properties, one expects that the macroscopic behavior is similar to that of an isotropic continuum. To check this intuitive conjecture, we endowed the geometric structure with two (simplest) types of mechanical interactions: the first one is a discrete-type model in which, along the edges of the tiling, we assume near-neighbor type pair interactions. As the tiling contains polygons, to avoid deformations of the polygons that do not change the lengths of their edges (zero modes) we also include three-body (angular) interactions in the simplest form of the Kirkwood-Keating approximation. The second type of mechanical structure is closer to a potential experiment: in this case, the edges of the tiling are regarded as beams endowed with extension/compression, shear and bending elastic energies. While the first mechanical model does not possess an (internal) length-scale, in the second one, the square root of the ratio between the second moment of area and the cross-sectional surface ($\sqrt{J/S}$) introduces an internal length-scale.   

In both cases, using linear Dirichlet boundary conditions, we compute the macroscopic response for an increasing sequence of length scales and several realizations at fixed length scale. The numerical results confirm a tendency toward a macroscopic isotropic behavior. In order to measure the proximity between a computed approximation of the fourth-order tensor of Hooke type and the subspace of isotropic fourth-order tensors of Hooke type, we introduced an anisotropy index defined in (\ref{I(H)}). The evolution of the set average of the anisotropy index as a function of the length scale shows a decrease by two orders of magnitude across the studied length scales (from $10a$ to $300a$). The reason for this (apparently slow) evolution is of geometric nature: the orientation of meta-tiles varies slowly with the inflation iteration index, and this variation is significantly slow for the monotile aperiodic hat tiling\footnote{Cf. our remark in Section 4.}.

In this regard, understanding the interplay between the geometric construction (rotation of meta-tiles during the inflation procedure) and the macroscopic behavior may be a key-step toward a proof of material isotropy at the macroscopic scale. This remains an important open question for future research.

%====================================================
% BIB file
%====================================================
\bibliographystyle{elsarticle-num} 
\bibliography{aperiodic.bib}

\end{document}

%% file: Figure3.tex
 \path[fill=gray!80] (   6.00,   1.73 ) -- (   4.50,   2.60 ) -- (   4.00,   1.73 ) -- (   4.50,   0.87 ) -- (   3.00,   0.00 ) -- (   3.00,  -1.73 ) -- (   4.00,  -1.73 ) -- (   4.50,  -2.60 ) -- (   6.00,  -1.73 ) -- (   7.50,  -2.60 ) -- (   8.00,  -1.73 ) -- (   7.00,   0.00 ) -- (   6.00,   0.00 ) --   cycle ;
 \path[fill=gray!80] (   8.00,  -5.20 ) -- (   7.50,  -4.33 ) -- (   9.00,  -3.46 ) -- (   9.00,  -1.73 ) -- (   8.00,  -1.73 ) -- (   7.50,  -2.60 ) -- (   6.00,  -1.73 ) -- (   4.50,  -2.60 ) -- (   5.00,  -3.46 ) -- (   4.50,  -4.33 ) -- (   6.00,  -5.20 ) -- (   6.00,  -6.93 ) -- (   7.00,  -6.93 ) --   cycle;
 \path[fill=gray!40]  (   2.00,  -5.20 ) -- (   1.50,  -6.06 ) -- (   3.00,  -6.93 ) -- (   4.50,  -6.06 ) -- (   5.00,  -6.93 ) -- (   6.00,  -6.93 ) -- (   6.00,  -5.20 ) -- (   4.50,  -4.33 ) -- (   5.00,  -3.46 ) -- (   4.00,  -1.73 ) -- (   3.00,  -1.73 ) -- (   3.00,  -3.46 ) -- (   1.50,  -4.33 ) --   cycle ; 
 
 \draw[red, line width = 0.25mm] (  -4.00,  -1.73 ) -- (  -4.50,  -0.87 ) -- (  -3.00,   0.00 ) -- (  -3.00,   1.73 ) -- (  -4.00,   1.73 ) -- (  -4.50,   0.87 ) -- (  -6.00,   1.73 ) -- (  -7.50,   0.87 ) -- (  -7.00,   0.00 ) -- (  -7.50,  -0.87 ) -- (  -6.00,  -1.73 ) -- (  -6.00,  -3.46 ) -- (  -5.00,  -3.46 ) --   cycle ;
 \draw[red, line width = 0.25mm] (  -3.00,  -6.93 ) -- (  -1.50,  -6.06 ) -- (  -2.00,  -5.20 ) -- (  -3.00,  -5.20 ) -- (  -3.00,  -3.46 ) -- (  -4.50,  -2.60 ) -- (  -5.00,  -3.46 ) -- (  -6.00,  -3.46 ) -- (  -6.00,  -5.20 ) -- (  -7.50,  -6.06 ) -- (  -7.00,  -6.93 ) -- (  -5.00,  -6.93 ) -- (  -4.50,  -6.06 ) --   cycle ;
 \draw[red, line width = 0.25mm] (  -2.00,  -5.20 ) -- (  -1.00,  -3.46 ) -- (  -1.50,  -2.60 ) -- (   0.00,  -1.73 ) -- (   0.00,   0.00 ) -- (  -1.00,   0.00 ) -- (  -1.50,  -0.87 ) -- (  -3.00,   0.00 ) -- (  -4.50,  -0.87 ) -- (  -4.00,  -1.73 ) -- (  -4.50,  -2.60 ) -- (  -3.00,  -3.46 ) -- (  -3.00,  -5.20 ) --   cycle ;
 \draw[red, line width = 0.25mm] (  -6.00,   5.20 ) -- (  -7.50,   6.06 ) -- (  -8.00,   5.20 ) -- (  -7.50,   4.33 ) -- (  -9.00,   3.46 ) -- (  -9.00,   1.73 ) -- (  -8.00,   1.73 ) -- (  -7.50,   0.87 ) -- (  -6.00,   1.73 ) -- (  -4.50,   0.87 ) -- (  -4.00,   1.73 ) -- (  -5.00,   3.46 ) -- (  -6.00,   3.46 ) --   cycle ;
 \draw[red, line width = 0.25mm] (   8.00,  -5.20 ) -- (   7.50,  -4.33 ) -- (   9.00,  -3.46 ) -- (   9.00,  -1.73 ) -- (   8.00,  -1.73 ) -- (   7.50,  -2.60 ) -- (   6.00,  -1.73 ) -- (   4.50,  -2.60 ) -- (   5.00,  -3.46 ) -- (   4.50,  -4.33 ) -- (   6.00,  -5.20 ) -- (   6.00,  -6.93 ) -- (   7.00,  -6.93 ) --   cycle ;
 \draw[red, line width = 0.25mm] (  -1.00,  -3.46 ) -- (  -1.50,  -4.33 ) -- (   0.00,  -5.20 ) -- (   0.00,  -6.93 ) -- (   1.00,  -6.93 ) -- (   2.00,  -5.20 ) -- (   1.50,  -4.33 ) -- (   3.00,  -3.46 ) -- (   3.00,  -1.73 ) -- (   2.00,  -1.73 ) -- (   1.50,  -2.60 ) -- (   0.00,  -1.73 ) -- (  -1.50,  -2.60 ) --   cycle ;
 \draw[red, line width = 0.25mm] (   2.00,  -5.20 ) -- (   1.50,  -6.06 ) -- (   3.00,  -6.93 ) -- (   4.50,  -6.06 ) -- (   5.00,  -6.93 ) -- (   6.00,  -6.93 ) -- (   6.00,  -5.20 ) -- (   4.50,  -4.33 ) -- (   5.00,  -3.46 ) -- (   4.00,  -1.73 ) -- (   3.00,  -1.73 ) -- (   3.00,  -3.46 ) -- (   1.50,  -4.33 ) --   cycle ;
 \draw[red, line width = 0.25mm] (   6.00,   1.73 ) -- (   4.50,   2.60 ) -- (   4.00,   1.73 ) -- (   4.50,   0.87 ) -- (   3.00,   0.00 ) -- (   3.00,  -1.73 ) -- (   4.00,  -1.73 ) -- (   4.50,  -2.60 ) -- (   6.00,  -1.73 ) -- (   7.50,  -2.60 ) -- (   8.00,  -1.73 ) -- (   7.00,   0.00 ) -- (   6.00,   0.00 ) --   cycle ;
 \draw[red, line width = 0.25mm] (   4.00,   1.73 ) -- (   2.00,   1.73 ) -- (   1.50,   0.87 ) -- (   0.00,   1.73 ) -- (  -1.50,   0.87 ) -- (  -1.00,   0.00 ) -- (   0.00,   0.00 ) -- (   0.00,  -1.73 ) -- (   1.50,  -2.60 ) -- (   2.00,  -1.73 ) -- (   3.00,  -1.73 ) -- (   3.00,   0.00 ) -- (   4.50,   0.87 ) --   cycle ;
 \draw[red, line width = 0.25mm] (   2.00,  -8.66 ) -- (   1.00,  -6.93 ) -- (   0.00,  -6.93 ) -- (   0.00,  -5.20 ) -- (  -1.50,  -4.33 ) -- (  -2.00,  -5.20 ) -- (  -1.50,  -6.06 ) -- (  -3.00,  -6.93 ) -- (  -3.00,  -8.66 ) -- (  -2.00,  -8.66 ) -- (  -1.50,  -9.53 ) -- (   0.00,  -8.66 ) -- (   1.50,  -9.53 ) --   cycle ;
 \draw[red, line width = 0.25mm] (   5.00,   6.93 ) -- (   4.00,   5.20 ) -- (   4.50,   4.33 ) -- (   3.00,   3.46 ) -- (   3.00,   1.73 ) -- (   4.00,   1.73 ) -- (   4.50,   2.60 ) -- (   6.00,   1.73 ) -- (   7.50,   2.60 ) -- (   7.00,   3.46 ) -- (   7.50,   4.33 ) -- (   6.00,   5.20 ) -- (   6.00,   6.93 ) --   cycle ;
 \draw[red, line width = 0.25mm] (   2.00,   1.73 ) -- (   3.00,   1.73 ) -- (   3.00,   3.46 ) -- (   4.50,   4.33 ) -- (   4.00,   5.20 ) -- (   2.00,   5.20 ) -- (   1.50,   4.33 ) -- (   0.00,   5.20 ) -- (  -1.50,   4.33 ) -- (  -1.00,   3.46 ) -- (   0.00,   3.46 ) -- (   0.00,   1.73 ) -- (   1.50,   0.87 ) --   cycle ;
 \draw[red, line width = 0.25mm] (   2.00,   5.20 ) -- (   3.00,   5.20 ) -- (   3.00,   6.93 ) -- (   1.50,   7.79 ) -- (   2.00,   8.66 ) -- (   1.50,   9.53 ) -- (   0.00,   8.66 ) -- (   0.00,   6.93 ) -- (  -1.00,   6.93 ) -- (  -2.00,   5.20 ) -- (  -1.50,   4.33 ) -- (   0.00,   5.20 ) -- (   1.50,   4.33 ) --   cycle ;
 \draw[red, line width = 0.25mm] (  -6.00,   5.20 ) -- (  -6.00,   3.46 ) -- (  -5.00,   3.46 ) -- (  -4.50,   4.33 ) -- (  -3.00,   3.46 ) -- (  -1.50,   4.33 ) -- (  -2.00,   5.20 ) -- (  -1.50,   6.06 ) -- (  -3.00,   6.93 ) -- (  -3.00,   8.66 ) -- (  -4.00,   8.66 ) -- (  -5.00,   6.93 ) -- (  -4.50,   6.06 ) --   cycle ;
 \draw[red, line width = 0.25mm] (  -5.00,   3.46 ) -- (  -4.00,   1.73 ) -- (  -3.00,   1.73 ) -- (  -3.00,   0.00 ) -- (  -1.50,  -0.87 ) -- (  -1.00,   0.00 ) -- (  -1.50,   0.87 ) -- (   0.00,   1.73 ) -- (   0.00,   3.46 ) -- (  -1.00,   3.46 ) -- (  -1.50,   4.33 ) -- (  -3.00,   3.46 ) -- (  -4.50,   4.33 ) --   cycle ;
 

%% file: IsoIndexNNAN.tex
%=====================================================
\begin{figure}[h!]
\centering
\begin{tikzpicture}[scale=1]
\begin{axis}[
xmin = 0, xmax = 320,ymin = 0.0, ymax = 0.1,
%ylabel={$\color{red} I({\mathbb H})$},ylabel near ticks,ylabel style={rotate=-90},
%y tick label style={/pgf/number format/zerofill, color=red}, 
scaled y ticks=false, grid=both,
xlabel={Domain diameter (in $a$ units)},
legend pos=north east,ymode = log]
\addplot[color=red,mark=*] 
      coordinates{

( 10.000000 , 0.076414 )
( 10.000000 , 0.091721 )
( 10.000000 , 0.065002 )
( 10.000000 , 0.020660 )
( 10.000000 , 0.090817 )
( 10.000000 , 0.078510 )
( 10.000000 , 0.049221 )
( 10.000000 , 0.070473 )
( 10.000000 , 0.105052 )
( 10.000000 , 0.087086 )

( 20.000000 , 0.033843 )
( 20.000000 , 0.024607 )
( 20.000000 , 0.020700 )
( 20.000000 , 0.041051 )
( 20.000000 , 0.035205 )
( 20.000000 , 0.021402 )
( 20.000000 , 0.028645 )
( 20.000000 , 0.027064 )
( 20.000000 , 0.017311 )
( 20.000000 , 0.013945 )

( 30.000000 , 0.015517 )
( 30.000000 , 0.009068 )
( 30.000000 , 0.015273 )
( 30.000000 , 0.008929 )
( 30.000000 , 0.016938 )
( 30.000000 , 0.005442 )
( 30.000000 , 0.015672 )
( 30.000000 , 0.004805 )
( 30.000000 , 0.011658 )
( 30.000000 , 0.015035 )

( 40.000000 , 0.008357 )
( 40.000000 , 0.008485 )
( 40.000000 , 0.010286 )
( 40.000000 , 0.007606 )
( 40.000000 , 0.004727 )
( 40.000000 , 0.008405 )
( 40.000000 , 0.007595 )
( 40.000000 , 0.006252 )
( 40.000000 , 0.004697 )
( 40.000000 , 0.002263 )

( 50.000000 , 0.006004 )
( 50.000000 , 0.001817 )
( 50.000000 , 0.005003 )
( 50.000000 , 0.006291 )
( 50.000000 , 0.004012 )
( 50.000000 , 0.009977 )
( 50.000000 , 0.004279 )
( 50.000000 , 0.005946 )
( 50.000000 , 0.005274 )
( 50.000000 , 0.003858 )

( 60.000000 , 0.004823 )
( 60.000000 , 0.003268 )
( 60.000000 , 0.002522 )
( 60.000000 , 0.004662 )
( 60.000000 , 0.004556 )
( 60.000000 , 0.003736 )
( 60.000000 , 0.003303 )
( 60.000000 , 0.005010 )
( 60.000000 , 0.003303 )
( 60.000000 , 0.006183 )

( 70.000000 , 0.005559 )
( 70.000000 , 0.002529 )
( 70.000000 , 0.002956 )
( 70.000000 , 0.003006 )
( 70.000000 , 0.003790 )
( 70.000000 , 0.002779 )
( 70.000000 , 0.003431 )
( 70.000000 , 0.003133 )
( 70.000000 , 0.003130 )
( 70.000000 , 0.003444 )

( 80.000000 , 0.001925 )
( 80.000000 , 0.003743 )
( 80.000000 , 0.002882 )
( 80.000000 , 0.002056 )
( 80.000000 , 0.002105 )
( 80.000000 , 0.002693 )
( 80.000000 , 0.003015 )
( 80.000000 , 0.003953 )
( 80.000000 , 0.001187 )
( 80.000000 , 0.002498 )

( 90.000000 , 0.002201 )
( 90.000000 , 0.003954 )
( 90.000000 , 0.002162 )
( 90.000000 , 0.004466 )
( 90.000000 , 0.001882 )
( 90.000000 , 0.002030 )
( 90.000000 , 0.003395 )
( 90.000000 , 0.001407 )
( 90.000000 , 0.002118 )
( 90.000000 , 0.001554 )

( 100.000000 , 0.002442 )
( 100.000000 , 0.002232 )
( 100.000000 , 0.002858 )
( 100.000000 , 0.002286 )
( 100.000000 , 0.001662 )
( 100.000000 , 0.001280 )
( 100.000000 , 0.002157 )
( 100.000000 , 0.002518 )
( 100.000000 , 0.002403 )
( 100.000000 , 0.001603 )      
      
( 120.000000 , 0.000803 )
( 120.000000 , 0.001123 )
( 120.000000 , 0.001070 )
( 120.000000 , 0.000599 )
( 120.000000 , 0.002087 )
( 120.000000 , 0.001551 )
( 120.000000 , 0.002198 )
( 120.000000 , 0.000870 )
( 120.000000 , 0.001136 )
( 120.000000 , 0.001614 )

( 140.000000 , 0.001387 )
( 140.000000 , 0.002070 )
( 140.000000 , 0.001348 )
( 140.000000 , 0.000950 )
( 140.000000 , 0.001972 )
( 140.000000 , 0.001310 )
( 140.000000 , 0.001357 )
( 140.000000 , 0.000873 )
( 140.000000 , 0.002074 )
( 140.000000 , 0.001188 )

( 160.000000 , 0.001406 )
( 160.000000 , 0.001547 )
( 160.000000 , 0.001094 )
( 160.000000 , 0.000616 )
( 160.000000 , 0.000575 )
( 160.000000 , 0.001092 )
( 160.000000 , 0.001891 )
( 160.000000 , 0.000832 )
( 160.000000 , 0.000745 )
( 160.000000 , 0.001113 )

( 180.000000 , 0.001290 )
( 180.000000 , 0.000477 )
( 180.000000 , 0.001121 )
( 180.000000 , 0.001170 )
( 180.000000 , 0.001165 )
( 180.000000 , 0.000202 )
( 180.000000 , 0.000924 )
( 180.000000 , 0.000482 )
( 180.000000 , 0.000738 )
( 180.000000 , 0.001176 )    

( 180.000000 , 0.001290 )
( 180.000000 , 0.000477 )
( 180.000000 , 0.001121 )
( 180.000000 , 0.001170 )
( 180.000000 , 0.001165 )
( 180.000000 , 0.000202 )
( 180.000000 , 0.000924 )
( 180.000000 , 0.000482 )
( 180.000000 , 0.000738 )
( 180.000000 , 0.001176 )

( 200.000000 , 0.000749 )
( 200.000000 , 0.000718 )
( 200.000000 , 0.000246 )
( 200.000000 , 0.000495 )
( 200.000000 , 0.001795 )
( 200.000000 , 0.000795 )
( 200.000000 , 0.001192 )
( 200.000000 , 0.000675 )
( 200.000000 , 0.001164 )
( 200.000000 , 0.000783 )

( 220.000000 , 0.000355 )
( 220.000000 , 0.000111 )
( 220.000000 , 0.000722 )
( 220.000000 , 0.001110 )
( 220.000000 , 0.000712 )
( 220.000000 , 0.000491 )
( 220.000000 , 0.000756 )
( 220.000000 , 0.000524 )
( 220.000000 , 0.000781 )
( 220.000000 , 0.000445 )

( 240.000000 , 0.000549 )
( 240.000000 , 0.000624 )
( 240.000000 , 0.000652 )
( 240.000000 , 0.000771 )
( 240.000000 , 0.000600 )
( 240.000000 , 0.000355 )
( 240.000000 , 0.000436 )
( 240.000000 , 0.000288 )
( 240.000000 , 0.000302 )
( 240.000000 , 0.000553 )

( 260.000000 , 0.000843 )
( 260.000000 , 0.000621 )
( 260.000000 , 0.000520 )
( 260.000000 , 0.000398 )
( 260.000000 , 0.000642 )
( 260.000000 , 0.000603 )
( 260.000000 , 0.000212 )
( 260.000000 , 0.000748 )
( 260.000000 , 0.000782 )
( 260.000000 , 0.001017 )

( 280.000000 , 0.000384 )
( 280.000000 , 0.000657 )
( 280.000000 , 0.000408 )
( 280.000000 , 0.000645 )
( 280.000000 , 0.000489 )
( 280.000000 , 0.000192 )
( 280.000000 , 0.000428 )
( 280.000000 , 0.000343 )
( 280.000000 , 0.000362 )
( 280.000000 , 0.000657 )

( 300.000000 , 0.000521 )
( 300.000000 , 0.000202 )
( 300.000000 , 0.000592 )
( 300.000000 , 0.000459 )
( 300.000000 , 0.000302 )
( 300.000000 , 0.000120 )
( 300.000000 , 0.000396 )
( 300.000000 , 0.000428 )
( 300.000000 , 0.000487 )
( 300.000000 , 0.000856 )
    
};

\addplot[color=black,mark=*] 
coordinates{    
( 10.000000 , 0.073496 )
( 20.000000 , 0.026377 )
( 30.000000 , 0.011834 )
( 40.000000 , 0.006867 )
( 50.000000 , 0.005246 )
( 60.000000 , 0.004137 )
( 70.000000 , 0.003376 )
( 80.000000 , 0.002606 )
( 90.000000 , 0.002517 )
( 100.000000 , 0.002144 )
( 100.000000 , 0.002144 )
( 120.000000 , 0.001305 )
( 140.000000 , 0.001453 )
( 160.000000 , 0.001091 )
( 180.000000 , 0.000874 )
( 180.000000 , 0.000874 )
( 200.000000 , 0.000861 )
( 220.000000 , 0.000601 )
( 240.000000 , 0.000513 )
( 260.000000 , 0.000639 )
( 280.000000 , 0.000456 )
( 300.000000 , 0.000436 )
};
    
\addlegendentry{$\color{red} I({\mathbb H})$}
\addlegendentry{$\color{black} \langle I({\mathbb H})\rangle$}
\end{axis}
\end{tikzpicture}
\caption{Evolution of the isotropy index $I({\mathbb H})$ as a function of length-scale on ten different realizations (in red); in black dots indicate the (set) average over ten realizations at all length-scales considered. 
The average anisotropy index is reduced by more than two orders of magnitude (from $4.8\cdot10^{-2}$ to $4.2\cdot10^{-4})$) a result that strongly indicates a convergence toward an isotropic material. Morevoer, at fixed the almost constant dispersion in the log-scale (at fixed $n$) indicates a realization independent result.}  
\label{Index}
\end{figure}

%% file: IsoIndexExtBend.tex
%=====================================================
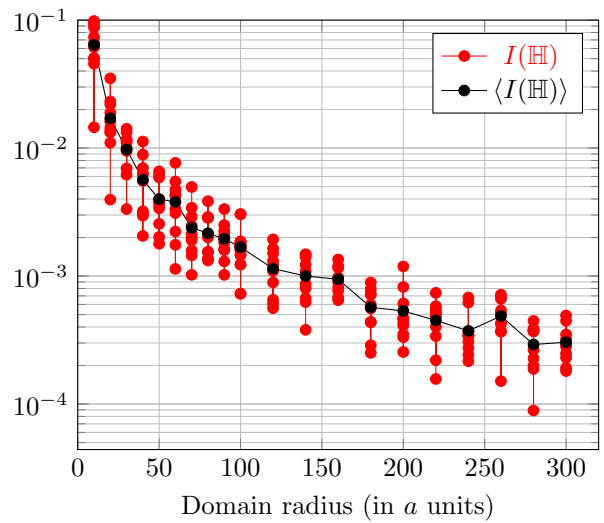
\begin{figure}[ht!]
\centering
\begin{tikzpicture}[scale=1]
\begin{axis}[
xmin = 0, xmax = 320,ymin = 0.0, ymax = 0.1,
%ylabel={$\color{red} I({\mathbb H})$},ylabel near ticks,ylabel style={rotate=-90},
%y tick label style={/pgf/number format/zerofill, color=red}, 
scaled y ticks=false, grid=both,
xlabel={Domain radius (in $a$ units)},
legend pos=north east,ymode = log]
\addplot[color=red,mark=*] 
      coordinates{
( 10.000000 , 0.087546 )
( 10.000000 , 0.098398 )
( 10.000000 , 0.061501 )
( 10.000000 , 0.014539 )
( 10.000000 , 0.073750 )
( 10.000000 , 0.092926 )
( 10.000000 , 0.049494 )
( 10.000000 , 0.066278 )
( 10.000000 , 0.045669 )
( 10.000000 , 0.050592 )

( 20.000000 , 0.014324 )
( 20.000000 , 0.023140 )
( 20.000000 , 0.013347 )
( 20.000000 , 0.035068 )
( 20.000000 , 0.018810 )
( 20.000000 , 0.011016 )
( 20.000000 , 0.021951 )
( 20.000000 , 0.016120 )
( 20.000000 , 0.013323 )
( 20.000000 , 0.003958 )

( 30.000000 , 0.011035 )
( 30.000000 , 0.003339 )
( 30.000000 , 0.014160 )
( 30.000000 , 0.013292 )
( 30.000000 , 0.006938 )
( 30.000000 , 0.006201 )
( 30.000000 , 0.011464 )
( 30.000000 , 0.009497 )
( 30.000000 , 0.010032 )
( 30.000000 , 0.011679 )

( 40.000000 , 0.006985 )
( 40.000000 , 0.011255 )
( 40.000000 , 0.008888 )
( 40.000000 , 0.003095 )
( 40.000000 , 0.005548 )
( 40.000000 , 0.002060 )
( 40.000000 , 0.006211 )
( 40.000000 , 0.002962 )
( 40.000000 , 0.006349 )
( 40.000000 , 0.003212 )

( 50.000000 , 0.001789 )
( 50.000000 , 0.003858 )
( 50.000000 , 0.003384 )
( 50.000000 , 0.005894 )
( 50.000000 , 0.003654 )
( 50.000000 , 0.006160 )
( 50.000000 , 0.003995 )
( 50.000000 , 0.006574 )
( 50.000000 , 0.002022 )
( 50.000000 , 0.002557 )

( 60.000000 , 0.007668 )
( 60.000000 , 0.001137 )
( 60.000000 , 0.001755 )
( 60.000000 , 0.003107 )
( 60.000000 , 0.002226 )
( 60.000000 , 0.004260 )
( 60.000000 , 0.003328 )
( 60.000000 , 0.004740 )
( 60.000000 , 0.005483 )
( 60.000000 , 0.004295 )

( 70.000000 , 0.004969 )
( 70.000000 , 0.001908 )
( 70.000000 , 0.002129 )
( 70.000000 , 0.002913 )
( 70.000000 , 0.001586 )
( 70.000000 , 0.001449 )
( 70.000000 , 0.001023 )
( 70.000000 , 0.003419 )
( 70.000000 , 0.001973 )
( 70.000000 , 0.002468 )

( 80.000000 , 0.002845 )
( 80.000000 , 0.001323 )
( 80.000000 , 0.002100 )
( 80.000000 , 0.001539 )
( 80.000000 , 0.001378 )
( 80.000000 , 0.002003 )
( 80.000000 , 0.003848 )
( 80.000000 , 0.002895 )
( 80.000000 , 0.001550 )
( 80.000000 , 0.002130 )

( 90.000000 , 0.001300 )
( 90.000000 , 0.002503 )
( 90.000000 , 0.002106 )
( 90.000000 , 0.003340 )
( 90.000000 , 0.001830 )
( 90.000000 , 0.001955 )
( 90.000000 , 0.001629 )
( 90.000000 , 0.002259 )
( 90.000000 , 0.001022 )
( 90.000000 , 0.001594 )

( 100.000000 , 0.001740 )
( 100.000000 , 0.001862 )
( 100.000000 , 0.003044 )
( 100.000000 , 0.001684 )
( 100.000000 , 0.001463 )
( 100.000000 , 0.001831 )
( 100.000000 , 0.001231 )
( 100.000000 , 0.000728 )
( 100.000000 , 0.001671 )
( 100.000000 , 0.001589 )

( 100.000000 , 0.001740 )
( 100.000000 , 0.001862 )
( 100.000000 , 0.003044 )
( 100.000000 , 0.001684 )
( 100.000000 , 0.001463 )
( 100.000000 , 0.001831 )
( 100.000000 , 0.001231 )
( 100.000000 , 0.000728 )
( 100.000000 , 0.001671 )
( 100.000000 , 0.001589 )

( 120.000000 , 0.001505 )
( 120.000000 , 0.001093 )
( 120.000000 , 0.001311 )
( 120.000000 , 0.000892 )
( 120.000000 , 0.001937 )
( 120.000000 , 0.000653 )
( 120.000000 , 0.001639 )
( 120.000000 , 0.000608 )
( 120.000000 , 0.001216 )
( 120.000000 , 0.000561 )

( 140.000000 , 0.000804 )
( 140.000000 , 0.001226 )
( 140.000000 , 0.001480 )
( 140.000000 , 0.000672 )
( 140.000000 , 0.001404 )
( 140.000000 , 0.001087 )
( 140.000000 , 0.000867 )
( 140.000000 , 0.000626 )
( 140.000000 , 0.001466 )
( 140.000000 , 0.000380 )

( 160.000000 , 0.000768 )
( 160.000000 , 0.001197 )
( 160.000000 , 0.000889 )
( 160.000000 , 0.000814 )
( 160.000000 , 0.000976 )
( 160.000000 , 0.000681 )
( 160.000000 , 0.001348 )
( 160.000000 , 0.000648 )
( 160.000000 , 0.000961 )
( 160.000000 , 0.001159 )

( 180.000000 , 0.000559 )
( 180.000000 , 0.000435 )
( 180.000000 , 0.000792 )
( 180.000000 , 0.000439 )
( 180.000000 , 0.000891 )
( 180.000000 , 0.000251 )
( 180.000000 , 0.000288 )
( 180.000000 , 0.000714 )
( 180.000000 , 0.000600 )
( 180.000000 , 0.000727 )

( 200.000000 , 0.000468 )
( 200.000000 , 0.000350 )
( 200.000000 , 0.000255 )
( 200.000000 , 0.000333 )
( 200.000000 , 0.001189 )
( 200.000000 , 0.000434 )
( 200.000000 , 0.000822 )
( 200.000000 , 0.000463 )
( 200.000000 , 0.000611 )
( 200.000000 , 0.000411 )

( 220.000000 , 0.000157 )
( 220.000000 , 0.000402 )
( 220.000000 , 0.000339 )
( 220.000000 , 0.000739 )
( 220.000000 , 0.000582 )
( 220.000000 , 0.000220 )
( 220.000000 , 0.000480 )
( 220.000000 , 0.000554 )
( 220.000000 , 0.000522 )
( 220.000000 , 0.000501 )

( 240.000000 , 0.000274 )
( 240.000000 , 0.000344 )
( 240.000000 , 0.000340 )
( 240.000000 , 0.000621 )
( 240.000000 , 0.000367 )
( 240.000000 , 0.000216 )
( 240.000000 , 0.000680 )
( 240.000000 , 0.000343 )
( 240.000000 , 0.000242 )
( 240.000000 , 0.000307 )

( 260.000000 , 0.000698 )
( 260.000000 , 0.000663 )
( 260.000000 , 0.000151 )
( 260.000000 , 0.000424 )
( 260.000000 , 0.000367 )
( 260.000000 , 0.000514 )
( 260.000000 , 0.000429 )
( 260.000000 , 0.000373 )
( 260.000000 , 0.000715 )
( 260.000000 , 0.000539 )

( 280.000000 , 0.000446 )
( 280.000000 , 0.000388 )
( 280.000000 , 0.000265 )
( 280.000000 , 0.000371 )
( 280.000000 , 0.000379 )
( 280.000000 , 0.000227 )
( 280.000000 , 0.000188 )
( 280.000000 , 0.000199 )
( 280.000000 , 0.000089 )
( 280.000000 , 0.000370 )

( 300.000000 , 0.000447 )
( 300.000000 , 0.000181 )
( 300.000000 , 0.000296 )
( 300.000000 , 0.000320 )
( 300.000000 , 0.000350 )
( 300.000000 , 0.000230 )
( 300.000000 , 0.000249 )
( 300.000000 , 0.000192 )
( 300.000000 , 0.000281 )
( 300.000000 , 0.000492 )

      };
\addplot[color=black,mark=*] 
coordinates{    
( 10.000000 , 0.064069 )
( 20.000000 , 0.017106 )
( 30.000000 , 0.009764 )
( 40.000000 , 0.005657 )
( 50.000000 , 0.003989 )
( 60.000000 , 0.003800 )
( 70.000000 , 0.002384 )
( 80.000000 , 0.002161 )
( 90.000000 , 0.001954 )
( 100.000000 , 0.001684 )
( 100.000000 , 0.001684 )
( 120.000000 , 0.001142 )
( 140.000000 , 0.001001 )
( 160.000000 , 0.000944 )
( 180.000000 , 0.000570 )
( 200.000000 , 0.000534 )
( 220.000000 , 0.000450 )
( 240.000000 , 0.000373 )
( 260.000000 , 0.000487 )
( 280.000000 , 0.000292 )
( 300.000000 , 0.000304 )

   };
    
\addlegendentry{$\color{red} I({\mathbb H})$}
\addlegendentry{$\color{black} \langle I({\mathbb H})\rangle$}
\end{axis}
\end{tikzpicture}
\caption{Evolution of the isotropy index $I({\mathbb H})$ as a function of length-scale on ten different realizations (in red) and the set-average over realizations (in black) at fixed length-scale. We notice that (i) the set-average value decreases (as expected) by two orders of magnitude when the size of the computation domain radius increases from $10a$ to $300a$ and (ii) the dispersion of the computed results of various realizations at fixed length-scale decrease with increasing length scale. Both of them indicate the macroscopic behavior of an isotropic material.}
\label{IndexRDM}
\end{figure}